\documentclass[amsmath,amssymb,longbibliography,floatfix,showpacs,notitlepage,nofootinbib,superscriptaddress,twocolumn]{revtex4-1}

\usepackage[T1]{fontenc}
\usepackage[utf8]{inputenc}
\usepackage{bm}
\usepackage{hyperref}
\usepackage{graphicx}
\usepackage{color}
\graphicspath{{oldfigures/}}
\hypersetup{colorlinks,
            citecolor=black,
            filecolor=black,
            linkcolor=black,
            urlcolor=blue }

\newcommand{\polb}{\bar{\mathbf{P}}}

\newcommand{\Sb}{\bar{S}}
\newcommand{\Pb}{\bar{P}}

\newcommand{\hv}{\protect{\hat{v}}}

\newcommand{\avg}[1]{\langle #1 \rangle}

\newcommand{\thetav}{\theta_\text{v}}
\newcommand{\weff}{\omega_\text{eff}}
\newcommand{\dd}{\protect{\text{d}}}
\newcommand{\ri}{\protect{\text{i}}}
\newcommand{\rv}{\protect{\vec{r}}}
\newcommand{\pv}{\protect{\vec{p}}}
\newcommand{\pol}{\mathbf{P}}

\newcommand{\Bv}{\mathbf{B}}
\newcommand{\Dv}{\mathbf{D}}

\newcommand{\Qv}{\mathbf{Q}}

\newcommand{\Vv}{\mathbf{V}}
\newcommand{\qv}{\mathbf{q}}
\newcommand{\be}{\mathbf{e}}
\newcommand{\GF}{G_{\text{F}}}
\newcommand{\eq}{\text{eq}}

\newcommand*{\UNM}{Department of Physics \& Astronomy, University of New
  Mexico, Albuquerque, NM 87131, USA}

\newcommand*{\APC}{Astro-Particule et Cosmologie (APC), CNRS UMR 7164,
  Universit\'e Denis Diderot, 75205 Paris Cedex 13, France}

\begin{document}

\title{Nonlinear flavor development of a two-dimensional neutrino gas}

\author{Joshua D. Martin}
\email{josh86@unm.edu}
\affiliation{\UNM}

\author{Sajad Abbar}
\affiliation{\UNM}
\affiliation{\APC}

\author{Huaiyu Duan}
\affiliation{\UNM}

\date{\today}

\begin{abstract}
  We present a numerical survey of the nonlinear flavor development of dense
  neutrino gases. This study is based on the stationary,
  two-dimensional ($x$ and $z$),
  two-beam, monochromatic neutrino line model with a periodic boundary condition
  along the $x$ direction. Similar to a previous work, we
  find that small-scale flavor structures can develop in a neutrino gas
  even if the physical conditions are nearly homogeneous along the $x$ axis
  initially. The power diffusion from the large-scale to small-scale
  structures increases with the neutrino density and helps to establish a
  semi-exponential dependence of the magnitudes of the Fourier moments
  on the corresponding wave numbers. The
  overall flavor conversion probabilities in the
  neutrino gases with small initial sinusoidal perturbations reach
  certain equilibrium values at large distances which are mainly determined by
  the neutrino-antineutrino asymmetry.
  Similar phenomena also exist in a neutrino gas with a
  localized initial perturbation, albeit only inside an expanding flavor
  conversion region. Our work suggests that a statistical treatment may
  be possible for the collective flavor oscillations of a dense
  neutrino gas in a multi-dimensional environment.
\end{abstract}

\maketitle

\section{Introduction} \label{sec:intro}

Due to the mismatch between its weak-interaction  and  vacuum
(mass) states, a neutrino can
experience flavor transformation or oscillations even in vacuum (see,
e.g., Ref.~\cite{Tanabashi:2018oca} for a review). In the limit of coherent
forward scattering
(i.e., with no momentum exchange), a neutrino can experience a
flavor dependent refraction in matter through the scattering by the
charged leptons and nucleons in ordinary matter
\cite{Wolfenstein:1977ue,Mikheev:1986gs}.
Similarly, the neutrinos in a dense neutrino medium
can also experience flavor dependent refraction through
the neutrino-neutrino scattering
\cite{Fuller:1987aa,Notzold:1987ik,Pantaleone:1992xh}. Unlike in
ordinary matter, however,
the so-called neutrino self-coupling potential can result in a
feature-rich, collective flavor transformation which is
coherent among different momentum modes of the neutrino medium (see,
e.g., Ref.~\cite{Duan:2010bg} for a review).

It is beyond the current means to solve the dynamic flavor
evolution in an dense, anistropic, inhomogeneous neutrino
medium. Instead, simplified models were studied with certain spatial
and directional symmetries imposed. For example, the isotropic and
homogeneous neutrino gas model has been employed to study the neutrino
oscillations in the early universe
(e.g., \cite{Kostelecky:1993yt,Abazajian:2002qx}), and the stationary,
spherical neutrino bulb model was used to study the neutrino flavor
transformation in supernovae (e.g., \cite{Duan:2006jv,Duan:2006an}).
The studies of these restricted models have brought many valuable
insights into this intriguing quantum phenomenon including the
flavor pendulum model \cite{Hannestad:2006nj} and the analytical
understanding of the spectral swaps/splits \cite{Raffelt:2007cb}.

The real physical environments such as that inside a
core-collapse supernova, however, can be highly anistropic and inhomogeneous.
Even if certain directional and spatial symmetries are approximately
held in a neutrino medium, they can be broken
spontaneously during the collective flavor transformation
\cite{Raffelt:2013rqa, Mirizzi:2013rla, Duan:2013kba, Mangano:2014zda,
  Duan:2014gfa, Chakraborty:2015tfa, Mirizzi:2015fva, Mirizzi:2015hwa, Capozzi:2016oyk}
(see also Ref.~\cite{Duan:2015cqa} for a review).
Consider, for example, the two-dimensional (2D) neutrino line model in which
the neutrinos are emitted from the ``neutrino line'' along the $x$ axis and
propagate in the $x$-$z$
plane. After linearizing the equation of motion one can show
\cite{Duan:2014gfa} that the amplitudes of the inhomogeneous oscillation
modes (whose flavor compositions vary along the $x$ axis) can grow
exponentially as functions of $z$ in the linear regime.
Unlike the restricted models, it is numerically challenging to
solve the neutrino oscillations without the spatial symmetries. As a
result, there have been only a few published works devoted to this subject
with very limited results
\cite{Mangano:2014zda,Mirizzi:2015fva, Mirizzi:2015hwa}.

In this work we presented a numerical survey on the inhomogeneous,
nonlinear flavor transformation of dense neutrino gases. This study is based
on the simplest 2D neutrino line model with two beams of monochromatic
neutrinos and antineutrinos emitted from each point along the neutrino
line. Through this work we wish to stimulate the interest of
developing an analytic understanding of the nonlinear flavor
transformation in the simplest 2D model which may bring
further insights into the more realistic models. To keep the problem
simple, we will ignore the temporal instabilities
\cite{Abbar:2015fwa, Dasgupta:2015iia}, fast flavor conversions
\cite{Sawyer:2015dsa, Chakraborty:2016lct}, the collisional effects
\cite{Cirigliano:2017hmk, Capozzi:2018clo}, and many other recent
developments (see Ref.~\cite{Chakraborty:2016yeg} for a review).

The rest of the paper is organized as follows. In
Sec.~\ref{sec:lineModel} we will describe the two-beam, neutrino line
model in detail and review the current analytic understanding of this
model.
In Sec.~\ref{sec:results} we will briefly describe the numerical
method we used to solve the line model and present a survey of the nonlinear
flavor transformation of the neutrino
gas in this model with various physical parameter choices.  In
Sec.~\ref{sec:conclusion} we give our conclusions.

\section{The neutrino line model} \label{sec:lineModel}
In this section we first establish the formalism by explaining the neutrino
line model proposed in Ref.~\cite{Duan:2014gfa}. We then briefly
review the flavor pendulum model \cite{Hannestad:2006nj} which
describes the flavor evolution in the neutrino line model
when both the spatial and directional symmetries are preserved. We
will also briefly
review the linearized flavor stability analysis performed in
Ref.~\cite{Duan:2014gfa} which describes the flavor evolution in the
linear regime.

\subsection{Equations of motion} \label{subsec:eom}

We study the mixing between two (effective) neutrino flavors $\nu_e$
and $\nu_\tau$, with $\nu_\tau$ being a suitable linear combination of
the physical $\nu_\mu$ and $\nu_\tau$. We use the (flavor) polarization vector
$\pol_\pv(t, \rv)$
in flavor space to represent the flavor quantum state of a neutrino of
momentum $\pv$ at time $t$ and position $\rv$.
We define the unit (flavor) basis vectors
$\be_i$ ($i=1,2,3$) such that $P_3=\pol\cdot\be_3$ gives the
probability $\mathcal{P}_{\nu_e}$ of the
neutrino being in the electron flavor through the relation
\begin{align}
  \mathcal{P}_{\nu_e} = \frac{1}{2}(1 + P_3),
\end{align}
and $P_1$ and $P_2$ describe the coherence between the two
flavors.  The polarization vector $\polb$ for the
antineutrino is defined in a similar way.

In absence of neutrino emission, absorption, and collision, the
polarization vectors obey the following equations of motion \cite{Sigl:1992fn}
\begin{subequations}
        \label{eq:eom-full}
\begin{align}
  (\partial_t + \hv\cdot\overrightarrow\nabla) \pol_{\pv}
  &= (\omega\Bv + \lambda \be_3 + \Vv_{\pv})\times \pol_{\pv},\\
  (\partial_t + \hv\cdot\overrightarrow\nabla) \polb_{\pv}
  &= (-\omega\Bv + \lambda \be_3 + \Vv_{\pv})\times \polb_{\pv}.
\end{align}
\end{subequations}
In the above equation, $\hv=\pv/|\pv|$ and $\omega=\delta m^2/2|\pv|$
are the velocity
and vacuum oscillation frequency of the neutrino, respectively,
$\Bv=\sin(2\thetav)\be_1 -\cos(2\thetav)\be_3$ is a unit vector
that denotes the neutrino mixing in vacuum, $\lambda=\sqrt2\GF n_e$ is
the matter potential with $\GF$ being the Fermi coupling
constant and  $n_e$  the net electron number density, and
\begin{widetext}
\begin{align}
  \Vv_\pv =
        \sqrt2\GF \int\!\frac{\dd^3 p'}{(2\pi)^3}
      (1-\hv\cdot\hv')\{[f_{\nu_e}(\pv') - f_{\nu_\tau}(\pv')]\pol_{\pv'}
        -[f_{\bar\nu_e}(\pv') - f_{\bar\nu_\tau}(\pv')]\polb_{\pv'}\},
\end{align}
\end{widetext}
is the neutrino self-coupling potential with
$f_\nu(\pv)$ ($\nu=\nu_e$, $\nu_\tau$,
$\bar\nu_e$ and $\bar\nu_\tau$) being the initial occupation numbers of
the neutrinos of momentum $\pv$ in the corresponding flavor states.

We adopt
the convention that the neutrino mass-squared difference $\delta m^2$
is always positive and that the normal and inverted neutrino mass
hierarchies (NH and IH) are represented by the vacuum mixing angles $\thetav$
within the ranges of $(0,\pi/4)$ and $(\pi/4,\pi/2)$, respectively.

We restrict our study to the two-beam, monchromatic neutrino line
model in which mono-energetic $\nu_e$'s and $\bar\nu_e$'s are
constantly emitted in two directions,
$\hv_\pm = (\pm u, 0, v_z)$, from each point on the $x$ axis,
where $u=\sqrt{1-v_z^2}$.
In this model, the convection operator becomes
\begin{align}
  \partial_t + \hv\cdot\overrightarrow\nabla \longrightarrow
  D_\pm = \pm u\partial_x + v_z\partial_z.
\end{align}

Because the presence of a large matter density does not
suppress collective oscillations in this model other than reducing the effective
neutrino mixing angle \cite{Duan:2005cp,Hannestad:2006nj}, we will
work in the appropriate frame rotating about $\be_3$ in flavor space
in which
\begin{align}
  \pm \omega\Bv
  + \lambda \be_3
  \xrightarrow{\lambda\gg\omega}
  \mp \eta\, \weff\, \be_3,
\end{align}
where $\eta=\text{sgn}(\cos2\thetav)$ is the signature of the neutrino
mass hierarchy, and
\begin{align}
  \weff=\omega\,|\cos2\thetav|
\end{align}
is the effective oscillation frequency of the neutrino.
In the rest of the paper, we will measure the energies in terms of
$\weff$ by setting
\begin{align}
  \weff=1.
\end{align}
Equation~\eqref{eq:eom-full} now reduces to
\begin{subequations}
  \label{eq:lineEOM}
\begin{align}
  D_\pm \pol_\pm &= [-\eta\be_3 +
  \mu(\pol_\mp - \alpha\polb_\mp) ] \times \pol_\pm, \\
  D_\pm \polb_\pm &= [+\eta\be_3 +
  \mu(\pol_\mp - \alpha\polb_\mp) ] \times \polb_\pm,
\end{align}
\end{subequations}
where the subscripts ``$\pm$'' denote the right- and left-going neutrino
beams, respectively. In Eq.~\eqref{eq:lineEOM}, the parameters
\begin{align}
  \mu = 2\sqrt2\GF n_{\nu_e} (1 - v_z^2)
\end{align}
and
\begin{align}
  \alpha = \frac{n_{\bar\nu_e}}{n_{\nu_e}}
\end{align}
measures the strength of the neutrino potential and the
neutrino-antineutrino asymmetry, respectively, where $n_{\nu_e}$
and $n_{\bar\nu_e}$ are the number densities of the corresponding
neutrino species.
We assume that both $\mu$ and $\alpha$ are constant in the whole space.

From Eq.~\eqref{eq:lineEOM} it is easy to show that the average
electron lepton number
\begin{align}
  \mathcal{L} = \frac{1}{4L}\int_0^L [(2+P_{3+} + P_{3-}) -
    \alpha(2+\Pb_{3+} + \Pb_{3-})]\,\dd x
  \label{eq:L}
\end{align}
is constant along $z$.

\subsection{The flavor pendulum} \label{subsec:bipolar}

If the neutrino emission is
homogeneous along the $x$ axis and symmetric between the two
directions, then
\begin{align}
  \pol_\pm(x, z) \longrightarrow \pol(z)
  \quad\text{and}\quad
  \polb_\pm(x, z) \longrightarrow \polb(z).
\end{align}
In this limit, the two-beam neutrino line model
reduces to the bipolar model \cite{Kostelecky:1994dt,Duan:2005cp}, and its
 flavor evolution is equivalent to the motion of a
pendulum in flavor space with a total angular
momentum $\Dv = \pol - \alpha\polb$ \cite{Hannestad:2006nj}.
Defining $\Qv = \pol + \alpha\polb +\eta\be_3/\mu$ and using
Eq.~\eqref{eq:lineEOM} one can show that
\begin{align}
  \dot\Dv =
  \qv \times (\eta |\Qv|) \be_3,
\end{align}
where $\dot F(z) = v_z\dd F/\dd z$ for an arbitrary function $F(z)$,
and $\qv=\Qv/|\Qv|$ and $(\eta|\Qv|)\be_3$ are the
``position'' and ``weight'' of the pendulum bob,
respectively. The spin $\sigma=\Dv\cdot\qv$ of the bob is conserved,
and
\begin{align}
  \frac{1}{\mu}\qv\times\dot\qv
  = \Dv - \sigma \qv
\end{align}
is the orbital angular momentum of the bob.

The pendulum model has been used to obtain
many useful insights of the flavor evolution of the bipolar model
\cite{Hannestad:2006nj,Duan:2007mv}. For example, one can easily see
that no (significant)
flavor transformation can occur at $\mu\gg1$ if the neutrino mass
hierarchy is normal ($\eta=+1$) because the flavor pendulum
is near its stable configuration initially. If $\eta=-1$ (IH),
flavor oscillation will occur
only if
\begin{align}
  \frac{2}{(1+\sqrt\alpha)^2}<\mu<\frac{2}{(1-\sqrt\alpha)^2},
  \label{eq:bipolar-regime}
\end{align}
and $P_3$ has a
minimum value
\begin{equation} \label{eq:p3Min}
  P_{3,\text{min}} = - \alpha + \frac{1}{4} \mu (1-\alpha)^{2} +
  \frac{1}{\mu}.
\end{equation}
No significant flavor oscillation occurs if $\mu>2/(1-\sqrt\alpha)^2$
where the flavor pendulum behaves like a sleeping top \cite{Duan:2007mv}.

\subsection{Spontaneous symmetry breaking} \label{subsec:linearAnalysis}

It turns out that an initially homogeneous neutrino gas can lose its
homogeneity during collective oscillations
\cite{Duan:2014gfa,Mirizzi:2015fva}. To see this,
we assume the following periodic boundary conditions along the $x$
axis:
\begin{align}
  \pol_\pm(0,z) = \pol_\pm(L, z)
  \quad\text{and}\quad
  \polb_\pm(0,z) = \polb_\pm(L, z)
\end{align}
as in Refs.~\cite{Duan:2014gfa,Mirizzi:2015fva},
where $L$ is the size of the periodic box. Performing the Fourier
transformation along the $x$ axis one obtains
\begin{align}
  \pol^{(m)}_\pm(z) = \frac{1}{L}\int_0^L e^{-\ri k_m x}
  \pol_\pm(x,z)\,\dd x
\end{align}
and a similar expression for $\polb^{(m)}_\pm(z)$,
where $k_m=2m\pi/L$ ($m=0,\pm1,\cdots$). Eq.~\eqref{eq:lineEOM} now
becomes
\begin{widetext}
  \begin{subequations}
    \label{eq:lineEOM-m}
\begin{align}
  \dot \pol^{(m)}_\pm
  &= \mp \ri k_m u\pol^{(m)}_\pm - \eta\be_3\times\pol^{(m)}_\pm
   +\mu\sum_{m'}[\pol^{(m')}_\mp -
     \alpha\polb^{(m')}_\mp]\times\pol^{(m-m')}_\pm \\
   \dot \polb^{(m)}_\pm
  &= \mp\ri k_m u\polb^{(m)}_\pm + \eta\be_3\times\polb^{(m)}_\pm
   +\mu\sum_{m'}[\pol^{(m')}_\mp - \alpha\polb^{(m')}_\mp]\times\polb^{(m-m')}_\pm.
\end{align}
\end{subequations}
\end{widetext}

Because of the coupling among the different Fourier moments, the
``power'' in the
low Fourier moments with small wave numbers $k_m$ can dissipate into
high moments and result in fine
structures and the loss of large-scale, coherent flavor evolution.
Before the translation symmetry is badly broken, however,
the Fourier moments of different wave numbers evolve independently.
This can be seen through the flavor stability analysis performed in
Ref.~\cite{Duan:2014gfa} which is recapped below.

In the linear regime where the neutrino coherence fields
\begin{subequations}
\begin{align}
  S^{(m)}_\pm &= \pol^{(m)}_\pm\cdot(\be_1 -\ri \be_2), \\
  \Sb^{(m)}_\pm &= \polb^{(m)}_\pm\cdot(\be_1 -\ri \be_2)
\end{align}
\end{subequations}
are small, they obey the linearized equations of motion:
\begin{subequations}
  \label{eq:linearEOM}
  \begin{align}
    \ri \dot S^{(m)}_\pm &=
        [-\eta\pm u k_m + (1-\alpha)\mu] S^{(m)}_\pm \nonumber\\
        &\quad- \mu[S^{(m)}_\mp - \alpha \Sb^{(m)}_\mp ], \\
    \ri \dot \Sb^{(m)}_\pm &=
        [+\eta\pm u k_m + (1-\alpha)\mu] \Sb^{(m)}_\pm \nonumber\\
        &\quad - \mu[S^{(m)}_\mp - \alpha \Sb^{(m)}_\mp ].
  \end{align}
\end{subequations}
The solution to the above equation is a linear superposition of the
normal modes:
\begin{align}
  \begin{bmatrix}
    S^{(m)}_+(z) \\ \Sb^{(m)}_+(z) \\
    S^{(m)}_-(z) \\ \Sb^{(m)}_-(z)
  \end{bmatrix}
  = \sum_{a=1}^4 \mathsf{S}^{(m)}_a e^{\ri K^{(m)}_a z},
\end{align}
where $\mathsf{S}^{(m)}_a$ and $K^{(m)}_a$ are the amplitudes and wave
numbers of the $a$th normal mode, respectively.
If there exist normal modes with $\text{Im}[K^{(m)}_a]>0$, the
corresponding Fourier moments are unstable against the flavor conversion,
and their amplitudes grow exponentially with $z$. In this case, even if
there is an approximate translation symmetry along the $x$ direction,
this symmetry is spontaneously broken when $|S^{(m\neq0)}|$ and
$|\Sb^{(m\neq0)}|$ grow to of $\mathcal{O}(1)$.

\begin{figure}[tbh!]
  \includegraphics[width=0.9\columnwidth]{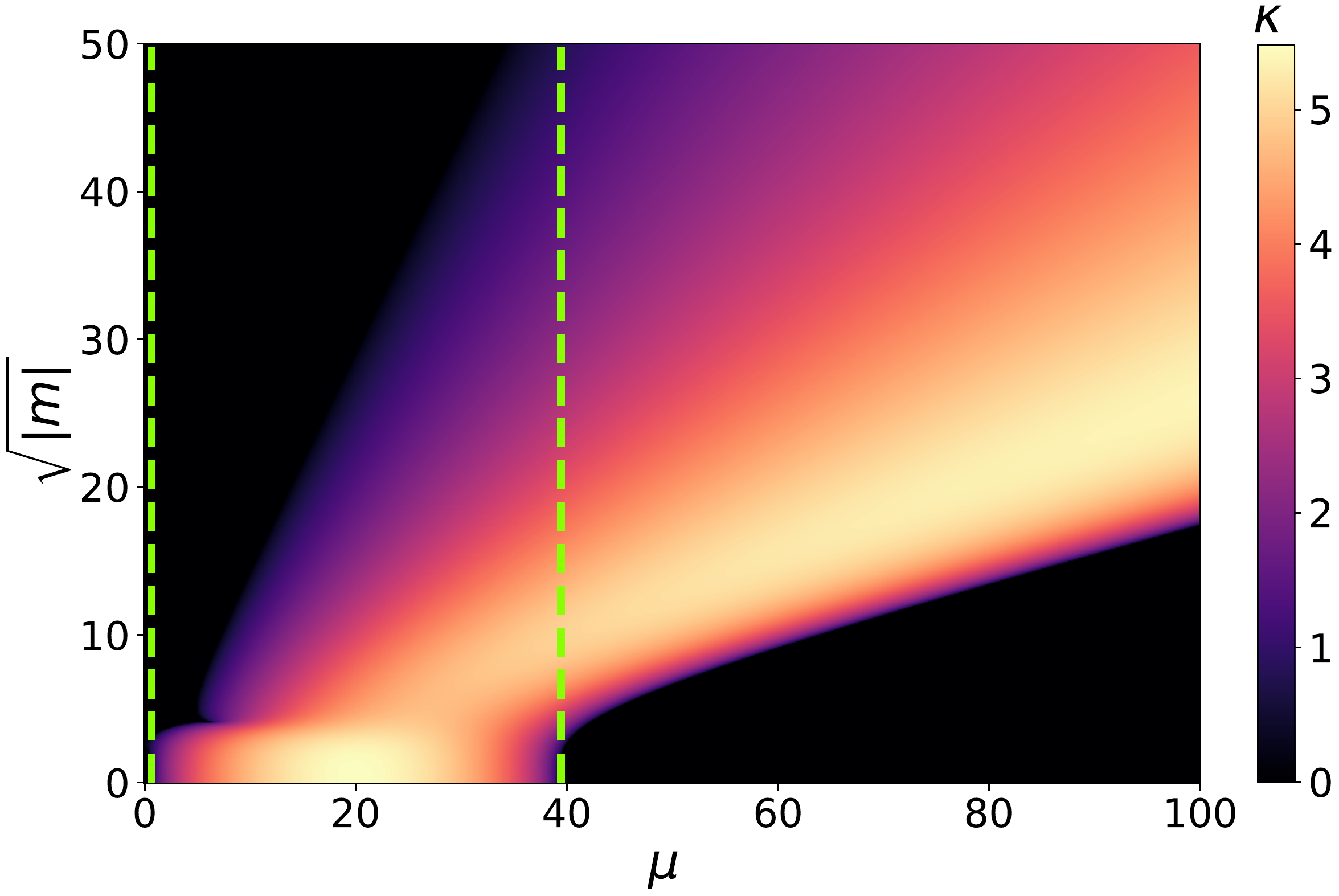}
  \caption{(Color online) The largest exponential growth rate
    $\kappa(m) = \text{max}(\text{Im}[K^{(m)}_a])$ of the flavor coherence
    amplitude as a function of the neutrino potential strength $\mu$
    and the Fourier moment index $m$
    in a two-beam neutrino line model with the
    neutrino-antineutrino asymetry $\alpha = 0.6$ and the
    neutrino velocity  $v_z=1/\sqrt2$ along the $z$ direction.
    The vertical dashed lines mark the boundaries of $\mu$ within which the
    homogeneous, bipolar model experiences a flavor oscillation like a
    flavor pendulum. This oscillation is suppressed at larger
    $\mu$ where the flavor pendulum behaves like a sleeping top.
  }
  \label{fig:kappa}
\end{figure}

In Fig.~\ref{fig:kappa}, we demonstrate the largest exponential growth
rate $\kappa(m) = \text{max}(\text{Im}[K^{(m)}_a])$ of the flavor coherence
amplitude as a function of both the strength of the neutrino potential
$\mu$ and the Fourier index $m$ for the two-beam neutrino line with
$\alpha=0.6$ and $v_z=1/\sqrt2$. We also marked the boundaries of the
regime where a homogeneous
neutrino gas is unstable against flavor
conversion [Eq.~\eqref{eq:bipolar-regime}].

\section{Numerical Results} \label{sec:results}

Following the pilot study in Ref.~\cite{Mirizzi:2015fva} which
features a single parameter set ($\eta=+1$, $\alpha\approx 0.77$, and
$\mu=13$), we have conducted a numerical survey of the two-beam
neutrino line model for a wide
range of the parameter space.%
We present a representative collection
of our numerical results in this section.
All the calculations assume the initial
conditions of the following form%
\begin{align}
  \pol_\pm(x,0) = \polb_\pm(x,0)
  &\approx [\epsilon_\pm, 0, 1],
  \label{eq:genIC}
\end{align}
where $|\epsilon_\pm(x)|\ll1$.
We choose the emission angles of both the
left- and right-going beams to be $\pi/4$. A change to the
neutrino emission angle is equivalent to a change of the neutrino
density and a rescaling of the $x$ and $z$ axes simultaneously.
We use a relatively large box with $L=30$ compared to $L\approx6.2$ in
Ref.~\cite{Mirizzi:2015fva}.

\newcommand{\figscalei}{0.3}
\begin{figure*}[htb]
  \begin{center}
    $\begin{array}{@{}l@{\hspace{0.01in}}l@{\hspace{0.01in}}l@{}}
      \includegraphics*[scale=\figscalei]{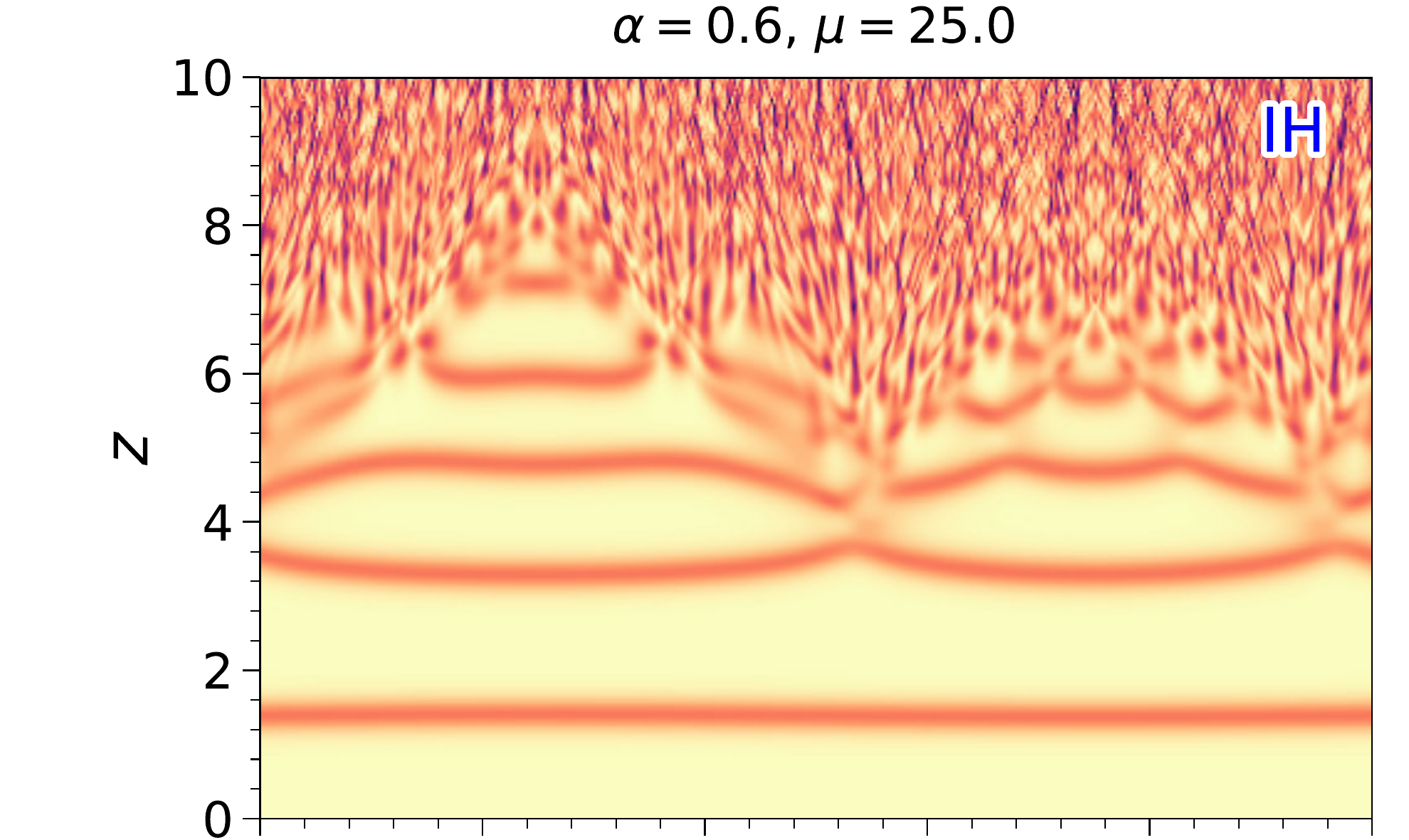} &
      \includegraphics*[scale=\figscalei]{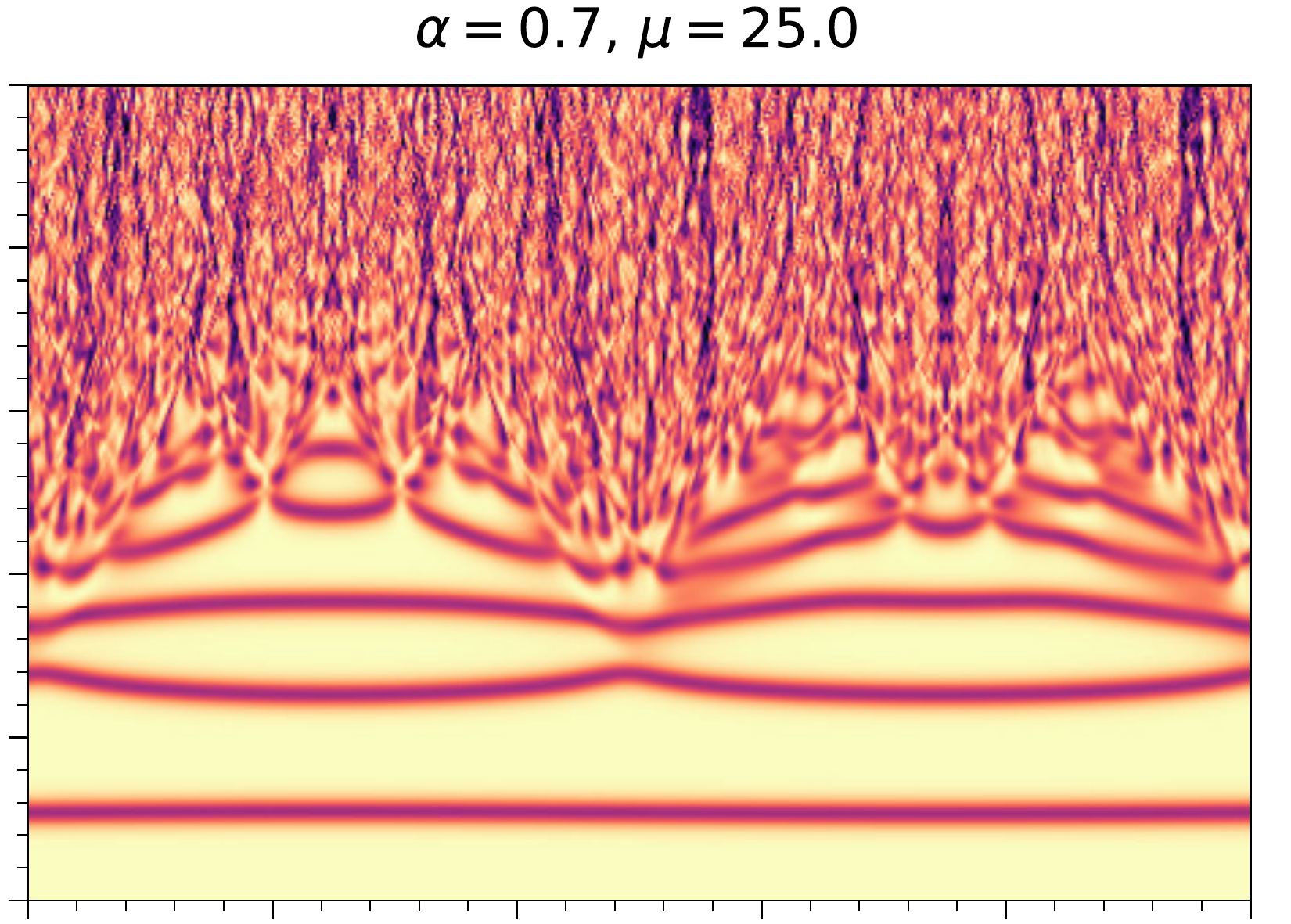} &
      \includegraphics*[scale=\figscalei]{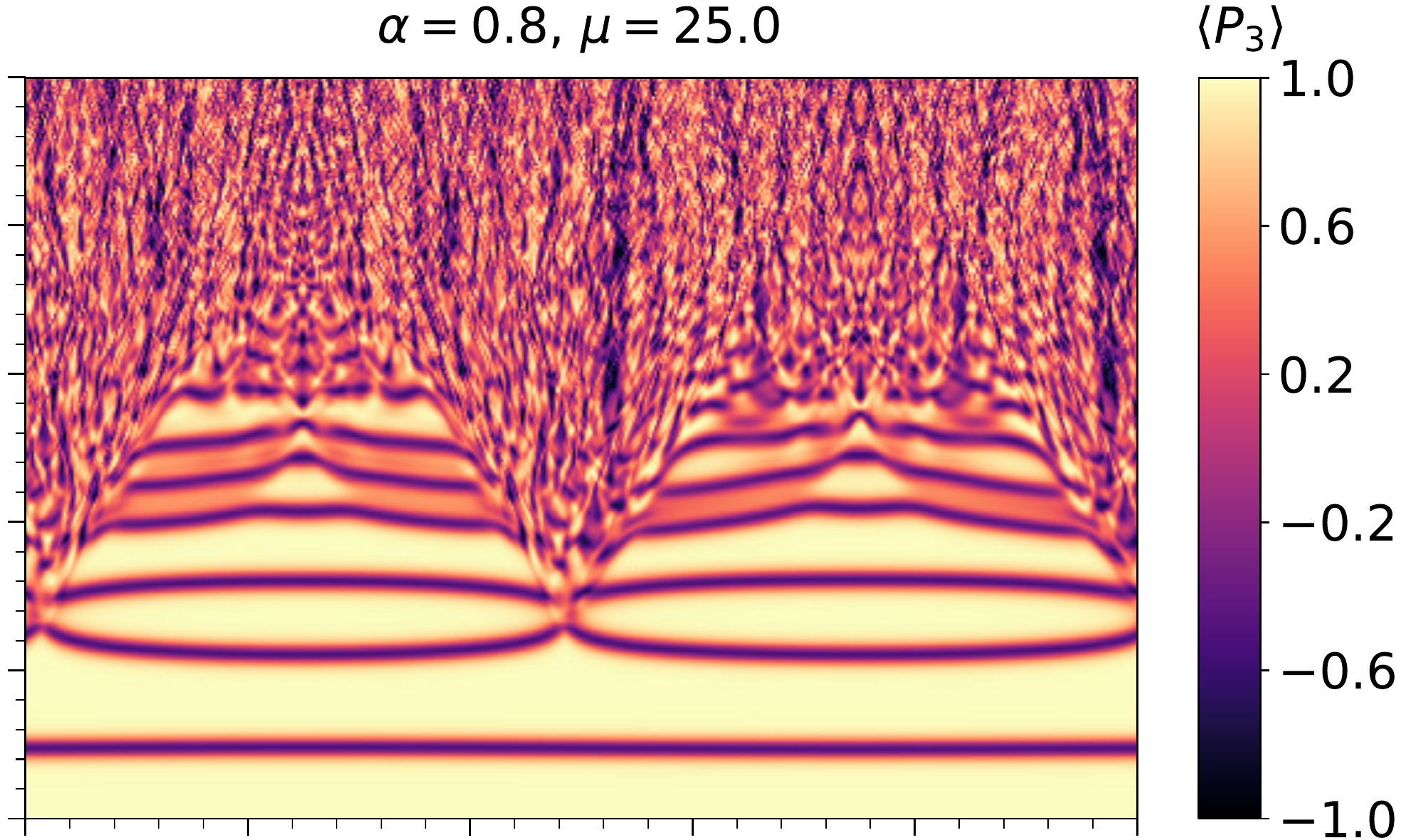} \\
      \includegraphics*[scale=\figscalei]{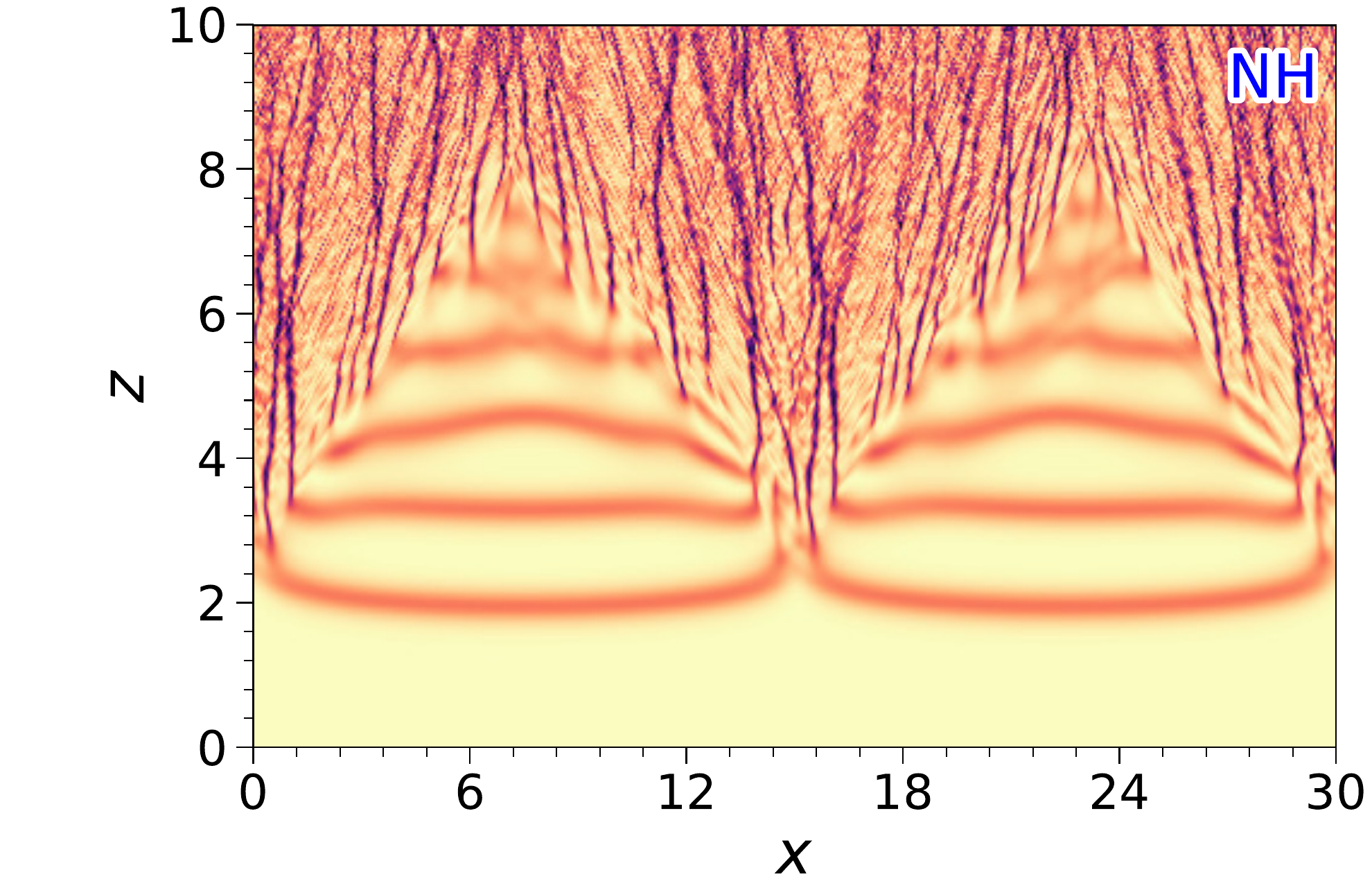} &
      \includegraphics*[scale=\figscalei]{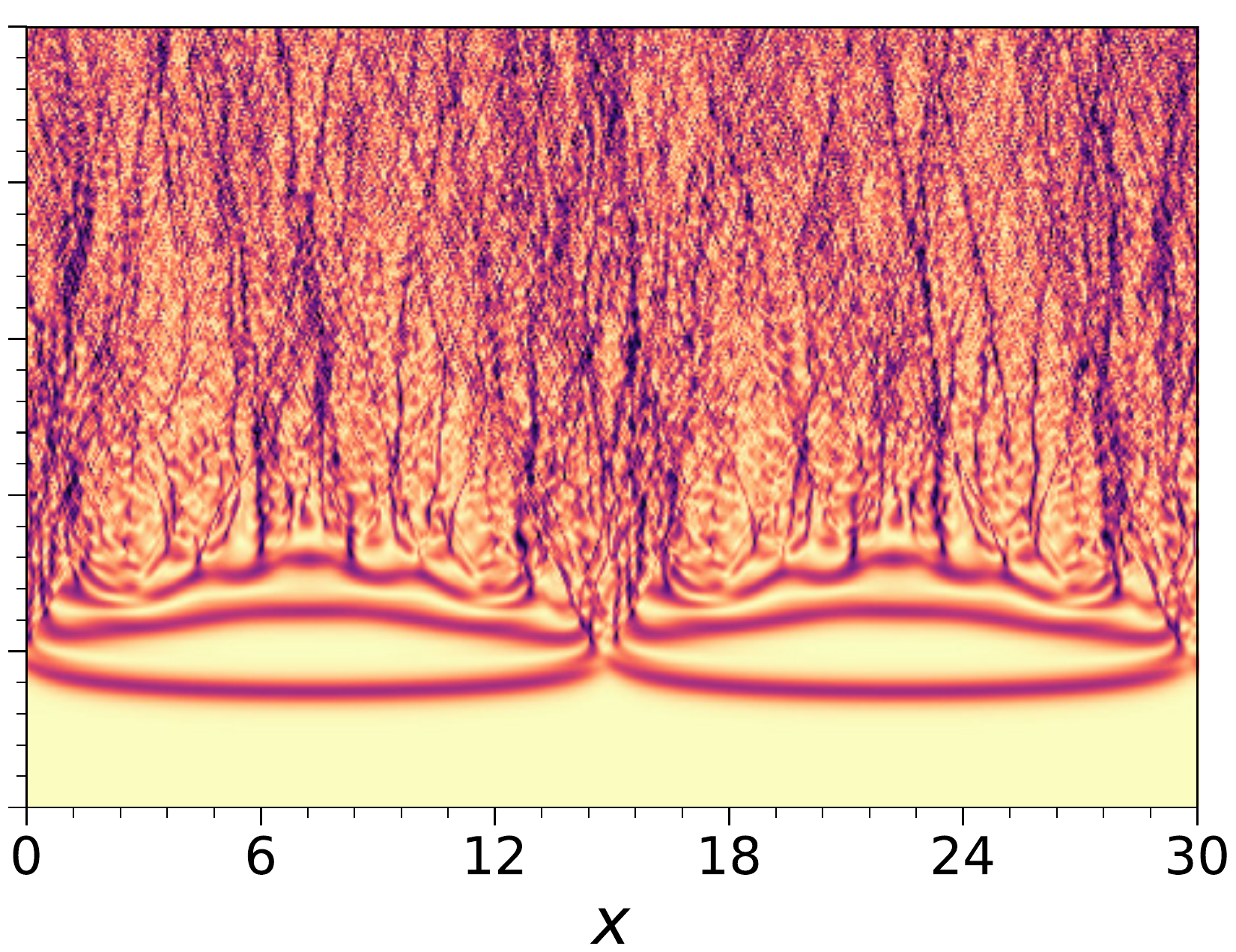} &
      \includegraphics*[scale=\figscalei]{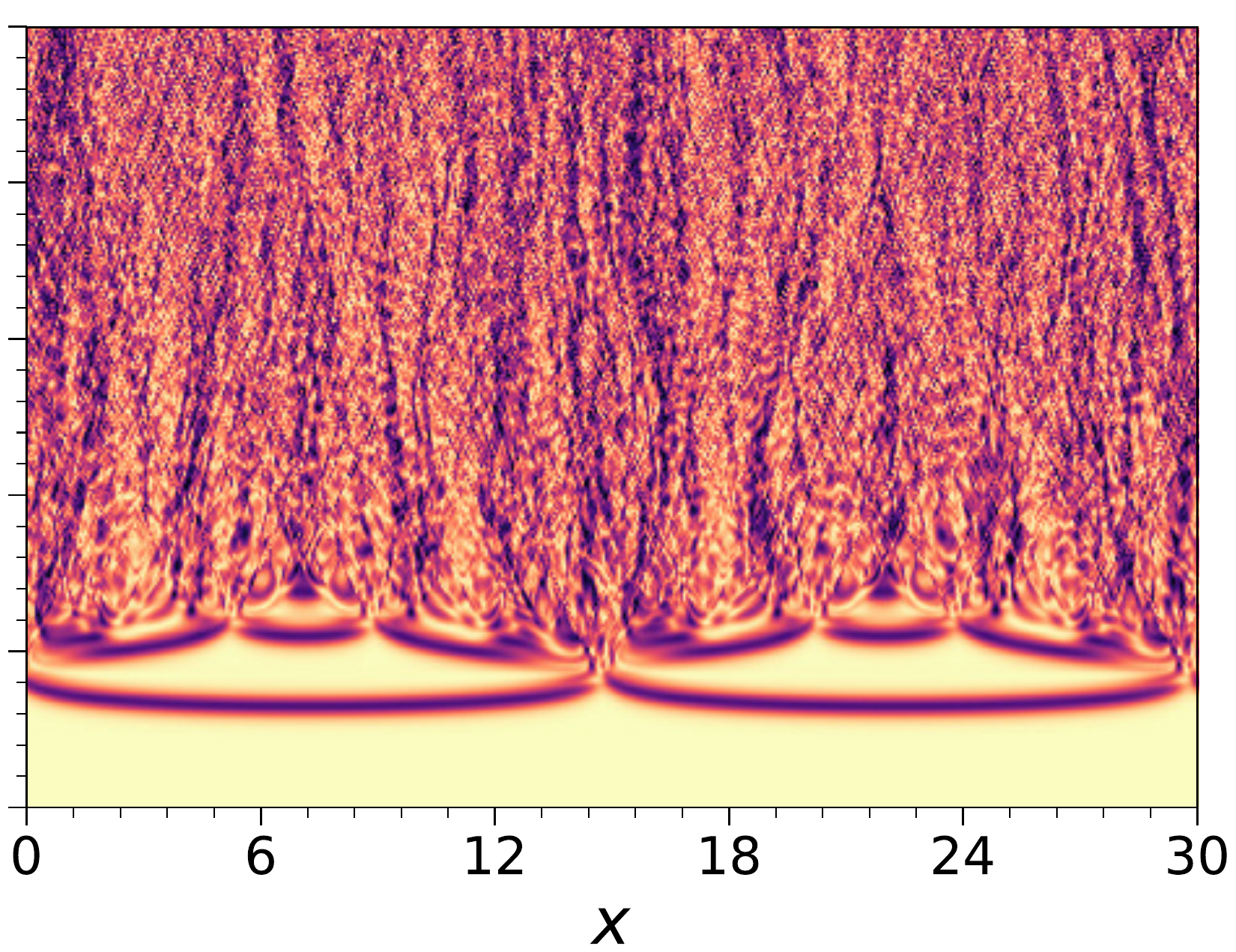} \\
      \includegraphics*[scale=\figscalei]{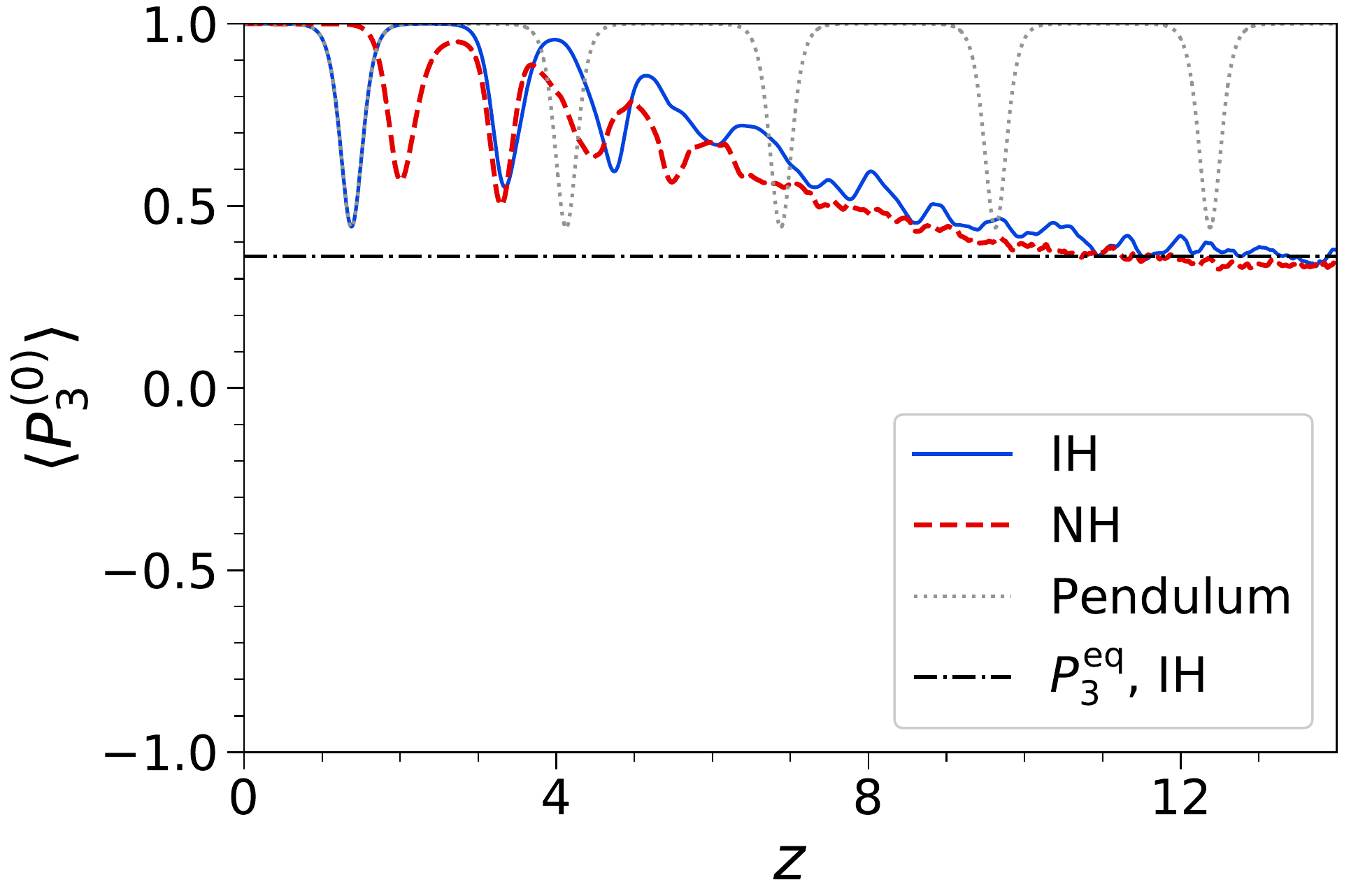} &
      \includegraphics*[scale=\figscalei]{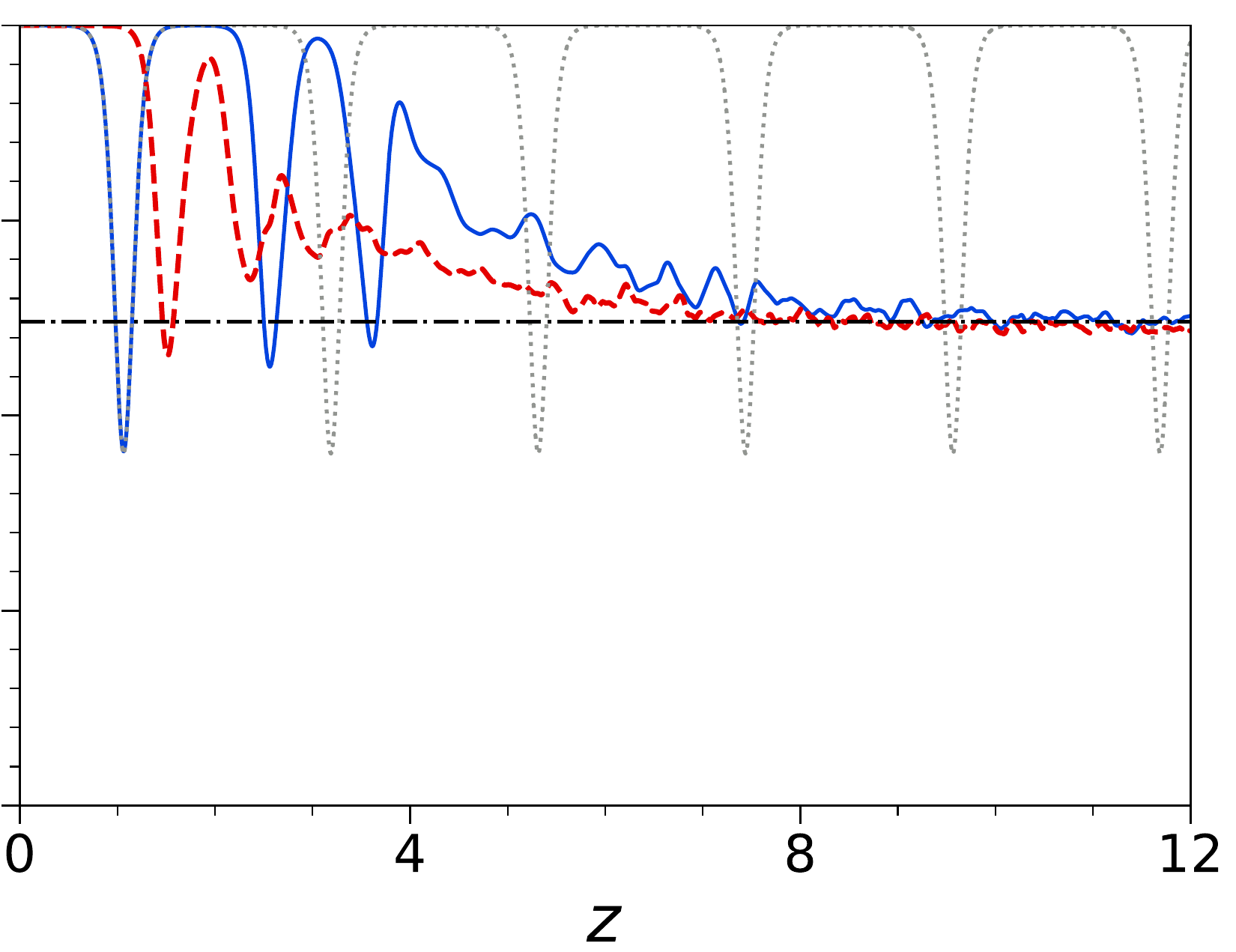} &
      \includegraphics*[scale=\figscalei]{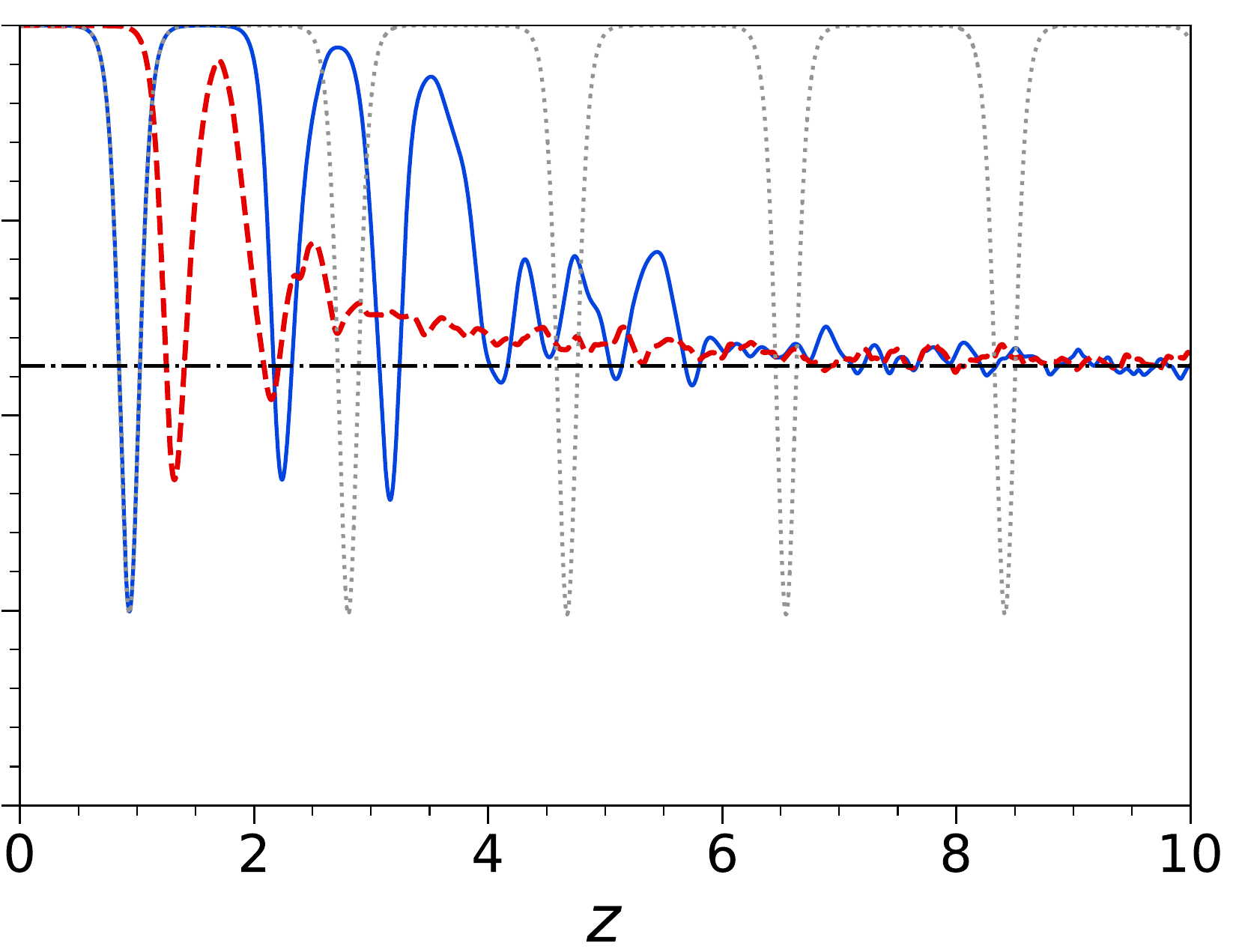}
    \end{array}$
  \end{center}
  \caption{(Color online)     \label{fig:P3-sine}
    The flavor evolution of the neutrino gas in the two-beam line
    model with the inverted (IH, top panels) and normal (NH, middle panels)
    neutrino mass hierarchies, various antineutrino-to-neutrino density
    ratios [$\alpha=0.6$ (left), $0.7$ (middle) and $0.8$ (right)],
    and sinusoidal initial perturbations described in the text. All
    the calculations have the same density of the neutrinos which is
    proportional to the neutrino self-coupling strength $\mu$. The
    top and middle panels show the angle-averaged polarization
    components $\avg{P_3}$ of
    the neutrinos as functions of $x$ and $z$, and the bottom panels
    show the evolution of these components  over $z$ when averaged over
    $x$. Also shown in the bottom panels are the values predicted by
    the pendulum model (dotted curves) and the equilibrium values
    $P^\eq_3$ in the IH calculations
    (dot-dashed lines).
  }
\end{figure*}

\subsection{Numerical method and validation} \label{subsec:algorithm}

Unlike Ref.~\cite{Mirizzi:2015fva} which solved $\pol_\pm^{(m)}(z)$ and
$\polb_\pm^{(m)}(z)$ from Eq.~\eqref{eq:lineEOM-m}, we solve
$\pol_\pm(x,z)$ and $\polb_\pm(x,z)$ in the direct space.
We discretize the $x$ axis into
$N$ equal intervals and solve the equations of motion by a finite
difference method derived from
the Lax-Wendroff algorithm \cite{NR2002}.
Our code accurately recovers the pendulum-like flavor oscillation when
$\epsilon_+=\epsilon_-=\text{const}$. In the linear regime where the
translation symmetry along the $x$ direction is slightly broken, our
code produces the correct exponential growth of the coherence
amplitudes as predicted by the linearized flavor stability analysis.
We varied
the number of discrete bins in the $x$ direction and the error tolerance
in the numerical integration along the $z$ direction to test the
numerical convergence of some representative calculations. We also varied the
size of the periodic box $L$ for the
calculations with localized perturbations to make sure that the results are
independent of the choice of $L$.
We do not enforce the unitary condition,
$|\pol(x,z)|=|\polb(x,z)|=1$, in our code but we rather use it to check the
correctness of the numerical solutions.

\subsection{Sinusoidal initial perturbations} \label{subsec:sinIC}

\newcommand{\figscaleii}{0.4}
\begin{figure*}[htb]
  \begin{center}
    $\begin{array}{@{}c@{\hspace{0.5in}}c@{}}
      \includegraphics*[scale=\figscaleii]{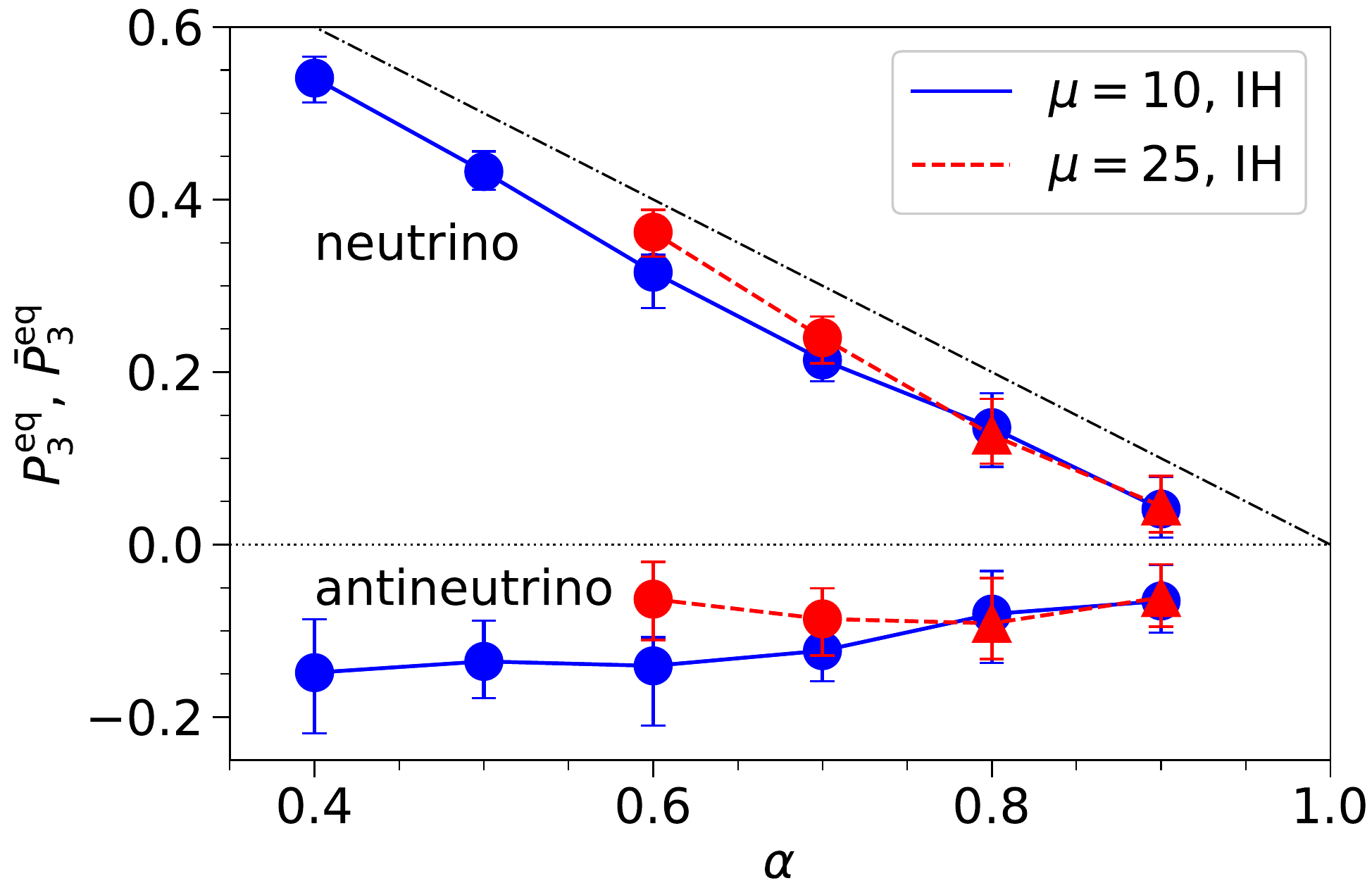} &
      \includegraphics*[scale=\figscaleii]{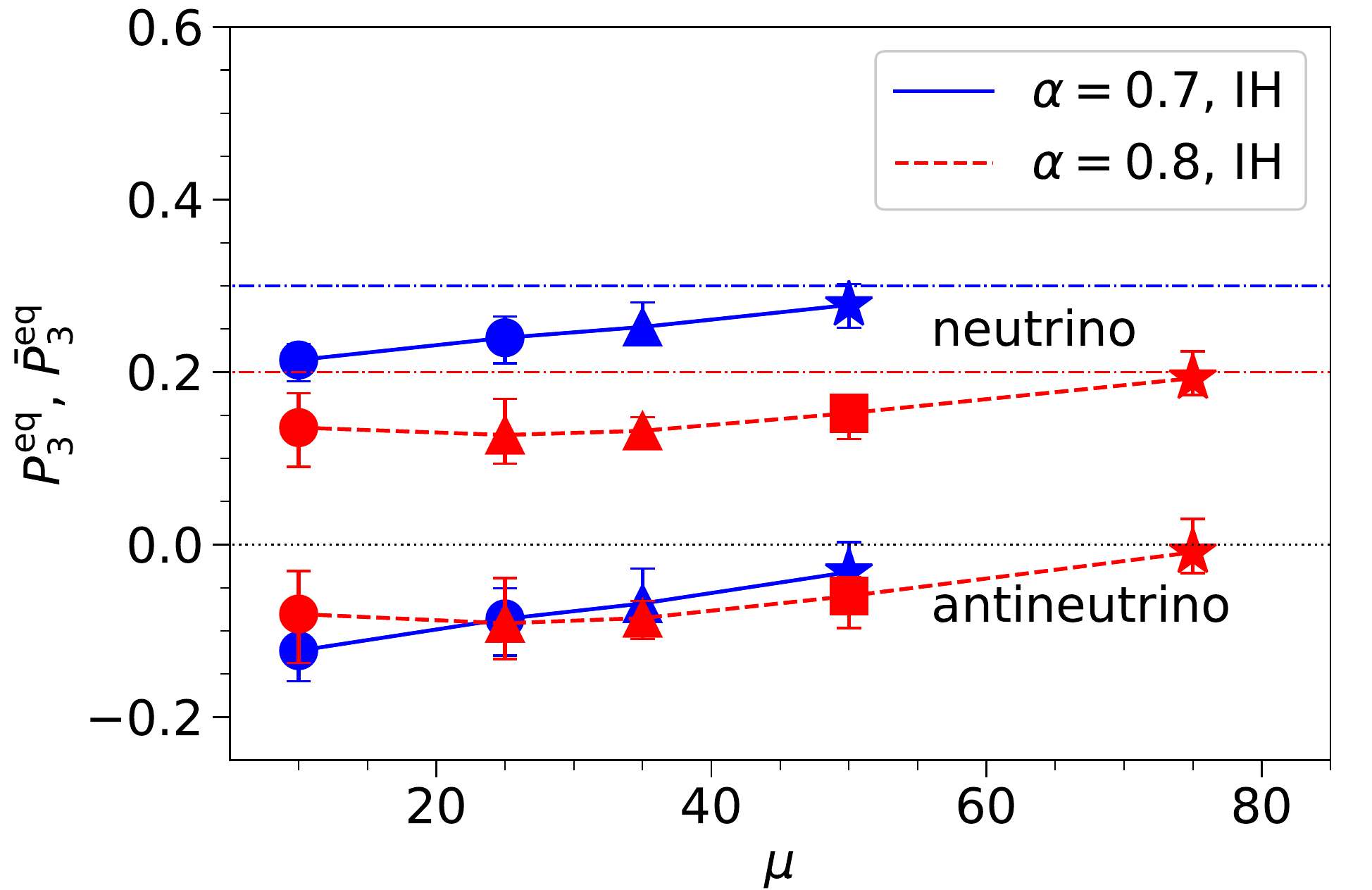}
    \end{array}$
  \end{center}
  \caption{(Color online) \label{fig:P3eq}
    The equilibrium values $P_3^\eq$ and $\Pb_3^\eq$ for the neutrinos
    (upper lines) and antineutrinos (lower lines) at large $z$ as
    functions of the antineutrino-to-neutrino ratio $\alpha$ (left) and
    as functions of the neutrino self-coupling strength $\mu$ (right),
    respectively. All the calculations
    assume the inverted neutrino mass hierarchy.
    The various symbols represent the calculations with $30,000$
    ({\Large$\bullet$}),
    $60,000$ ($\blacktriangle$), $120,000$ ({\footnotesize$\blacksquare$}), and
    $240,000$ ($\bigstar$) discrete $x$ bins, respectively. The
    error bars indicate the maximal and minimal values of
    $\avg{P_3^{(0)}}$ and $\avg{\Pb_3^{(0)}}$
    in the distance range over which the (mean) equilibrium values are
    calculated. The dotted lines represent the case with $\Pb_3^\eq=0$
    where the antineutrinos are fully depolarized in flavor, and the
    dot-dashed lines are the corresponding values of $P_3^\eq$
    obtained from Eq.~\eqref{eq:P3eq-cons}.
  }
\end{figure*}

\begin{figure*}[htb]
  \begin{center}
    $\begin{array}{@{}l@{\hspace{0.01in}}l@{\hspace{0.01in}}l@{}}
      \includegraphics*[scale=\figscalei]{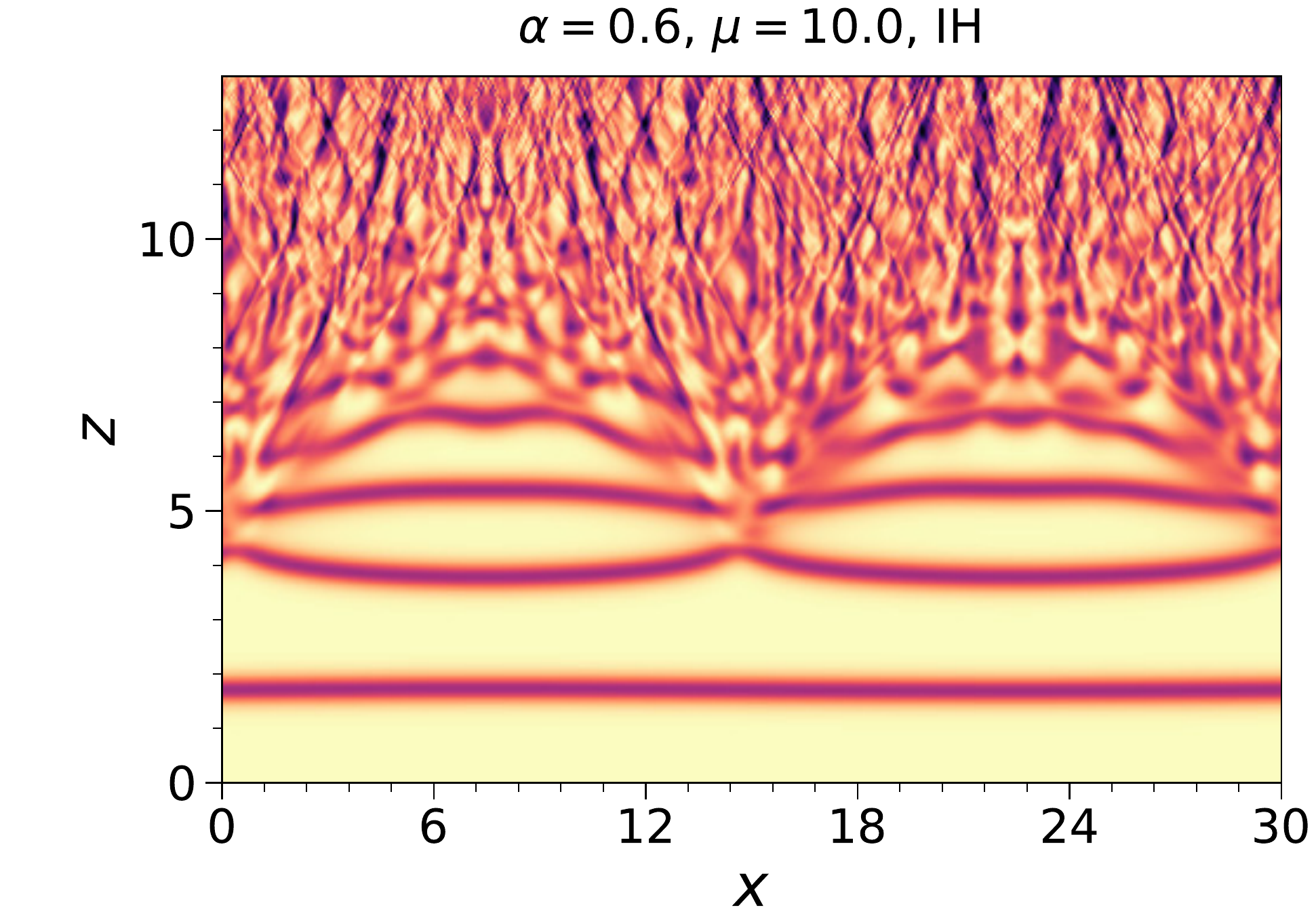} &
      \includegraphics*[scale=\figscalei]{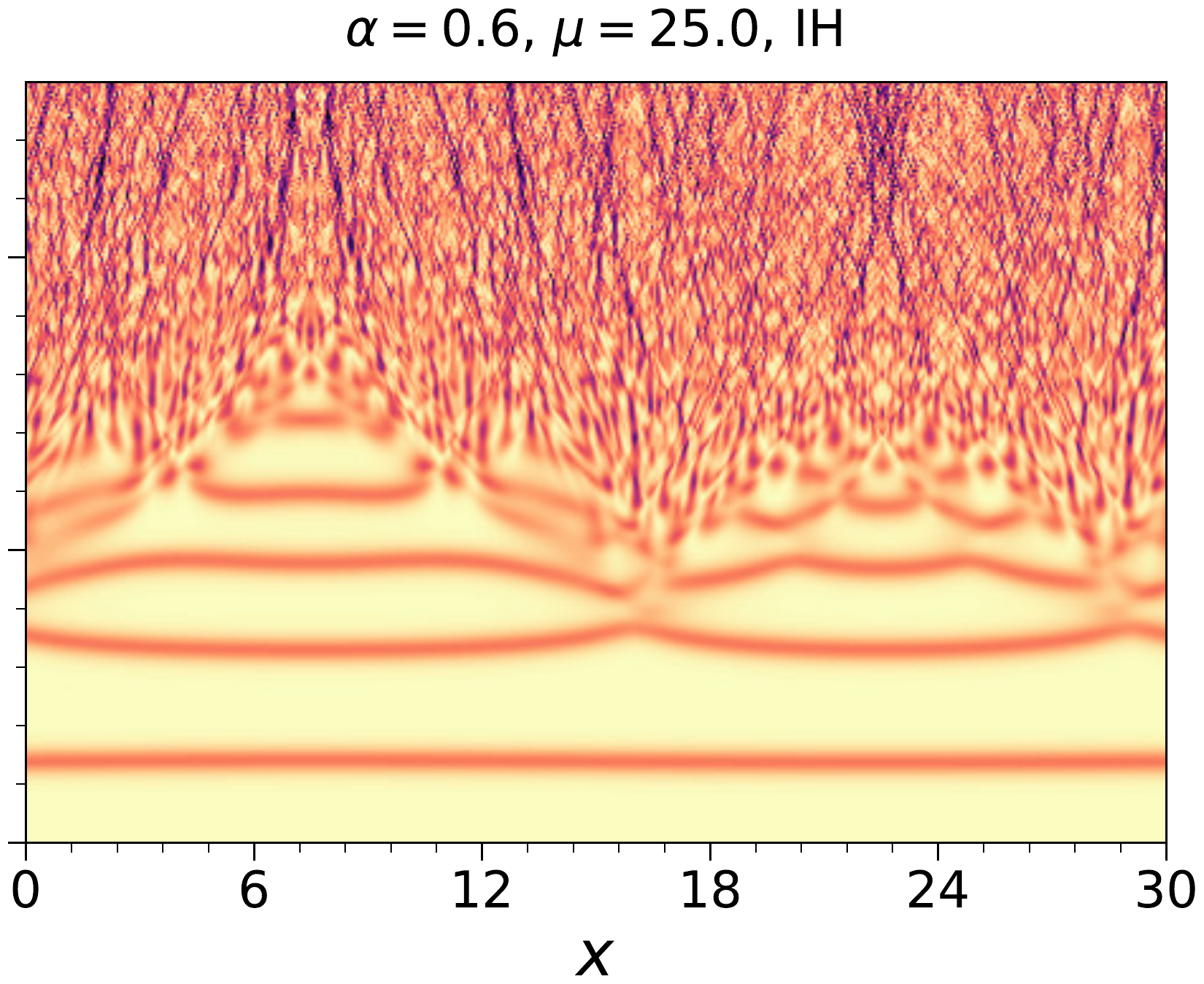} &
      \includegraphics*[scale=\figscalei]{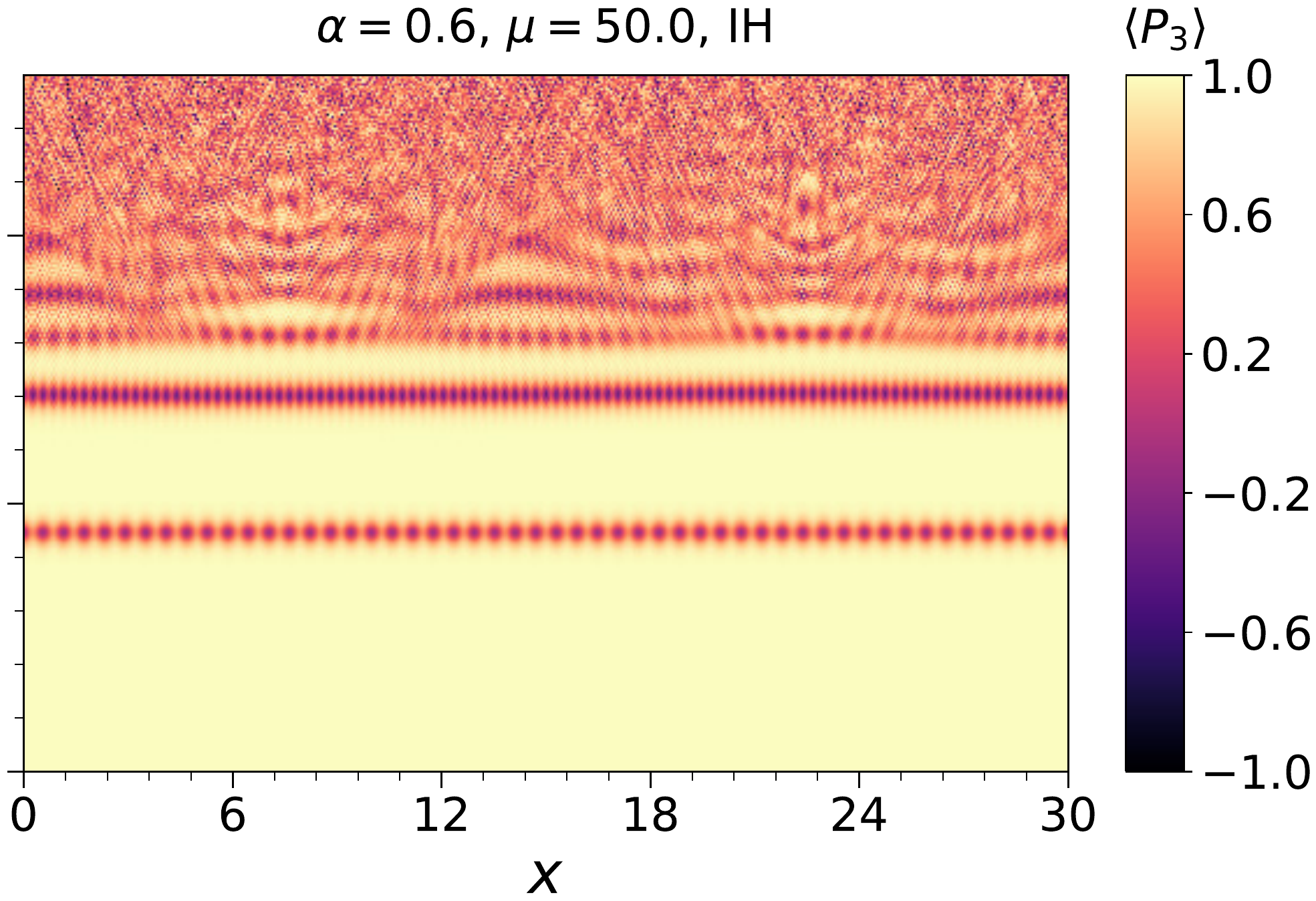} \\
      \includegraphics*[scale=\figscalei]{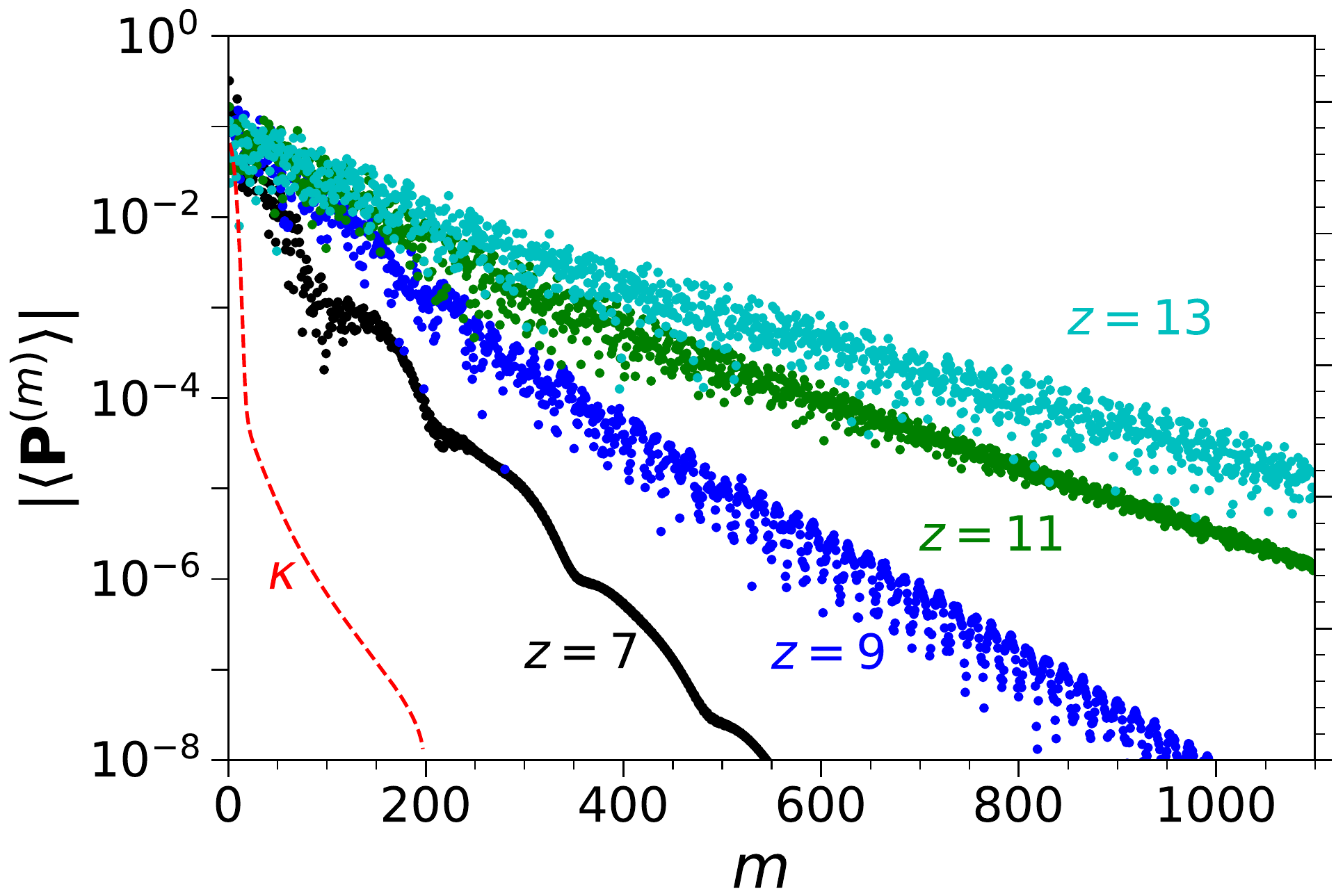} &
      \includegraphics*[scale=\figscalei]{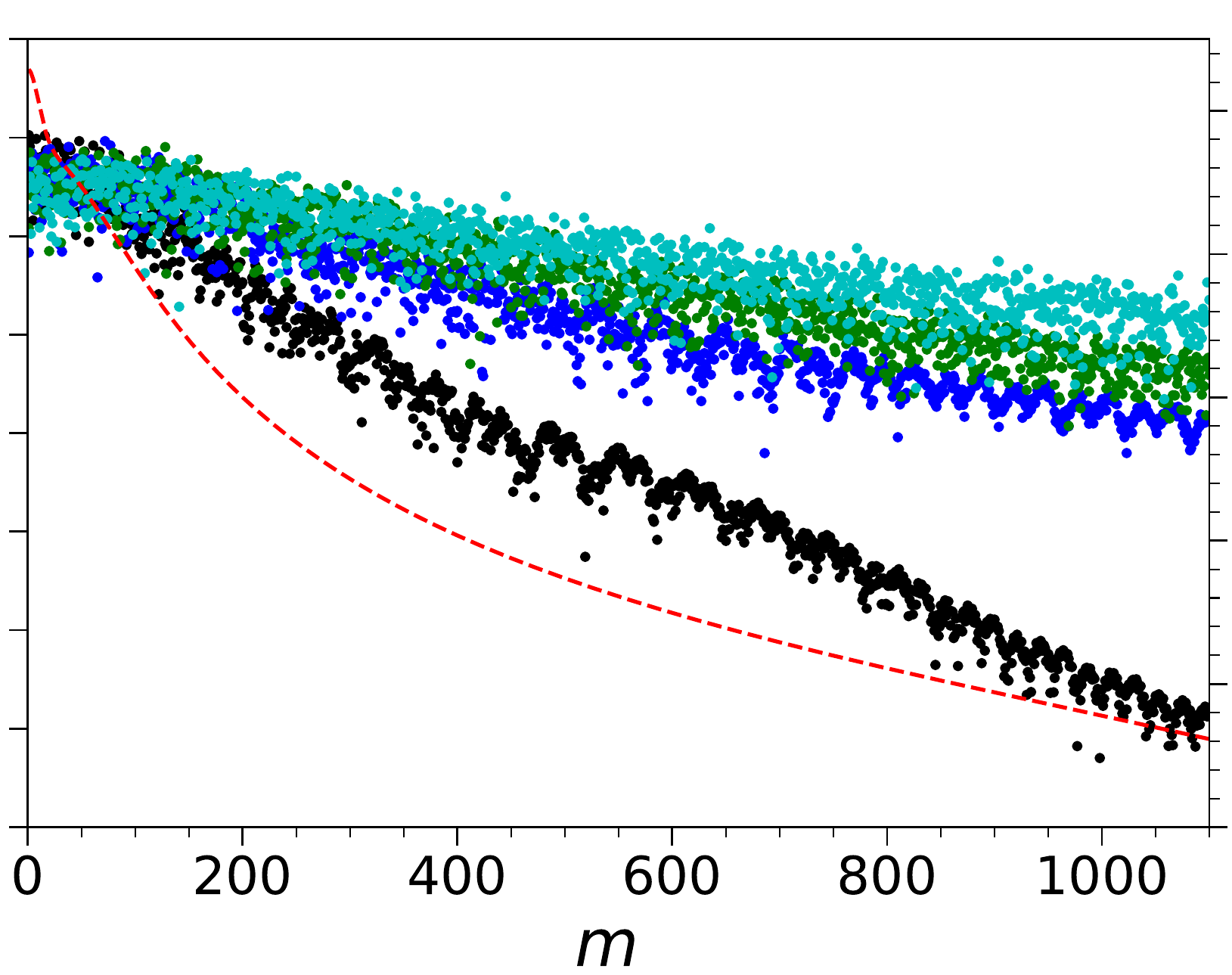} &
      \includegraphics*[scale=\figscalei]{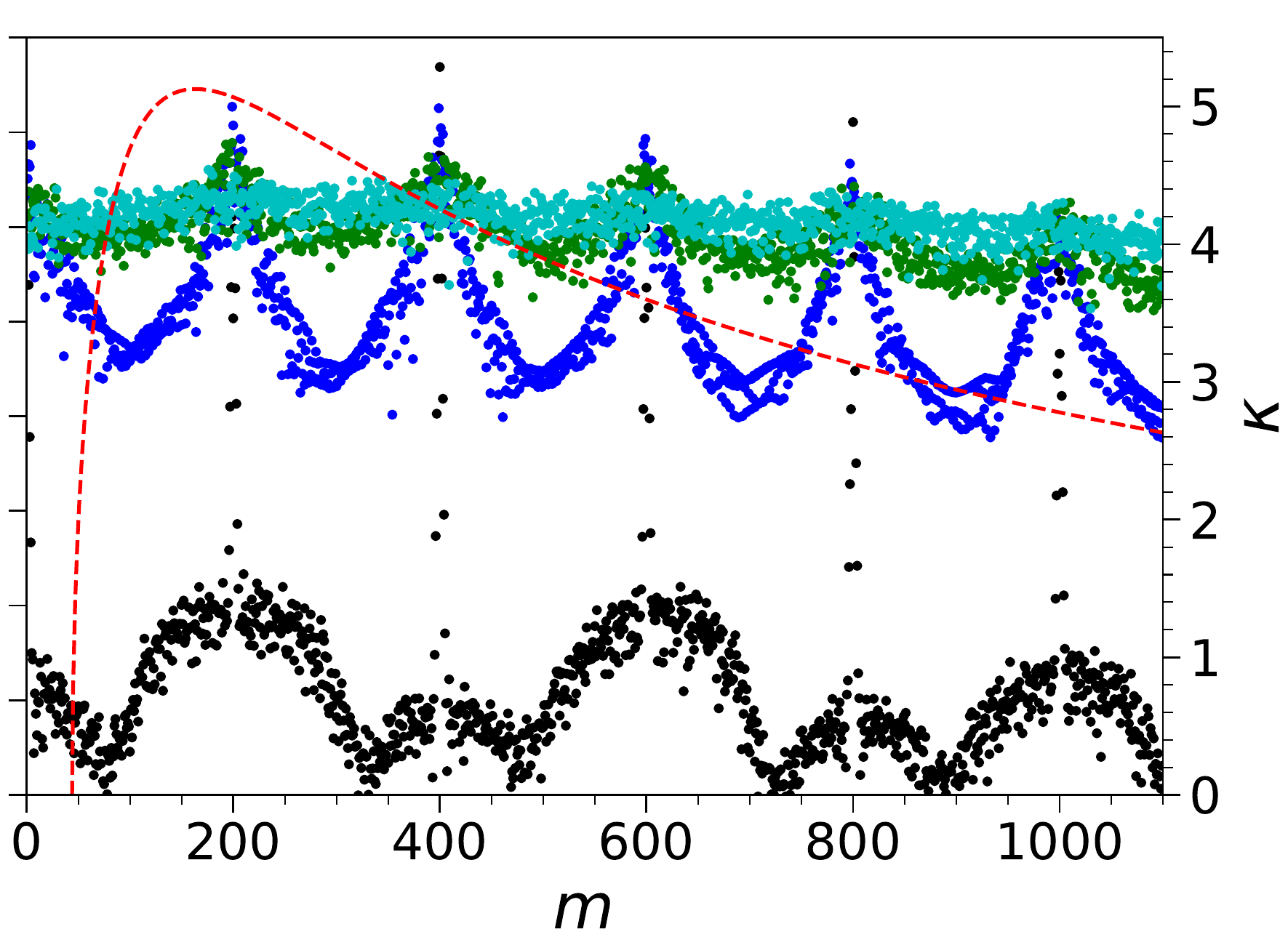} \\
      \includegraphics*[scale=\figscalei]{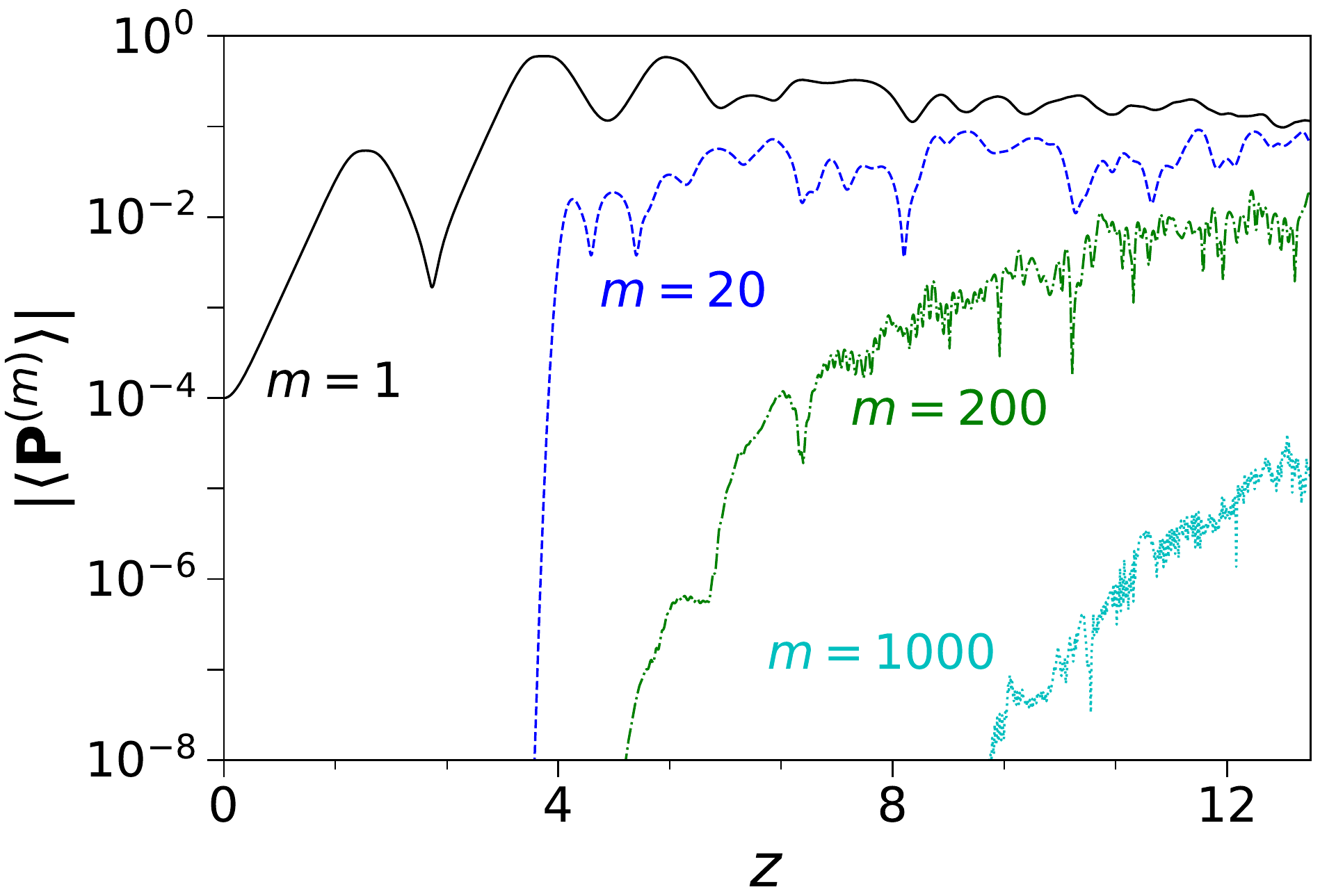} &
      \includegraphics*[scale=\figscalei]{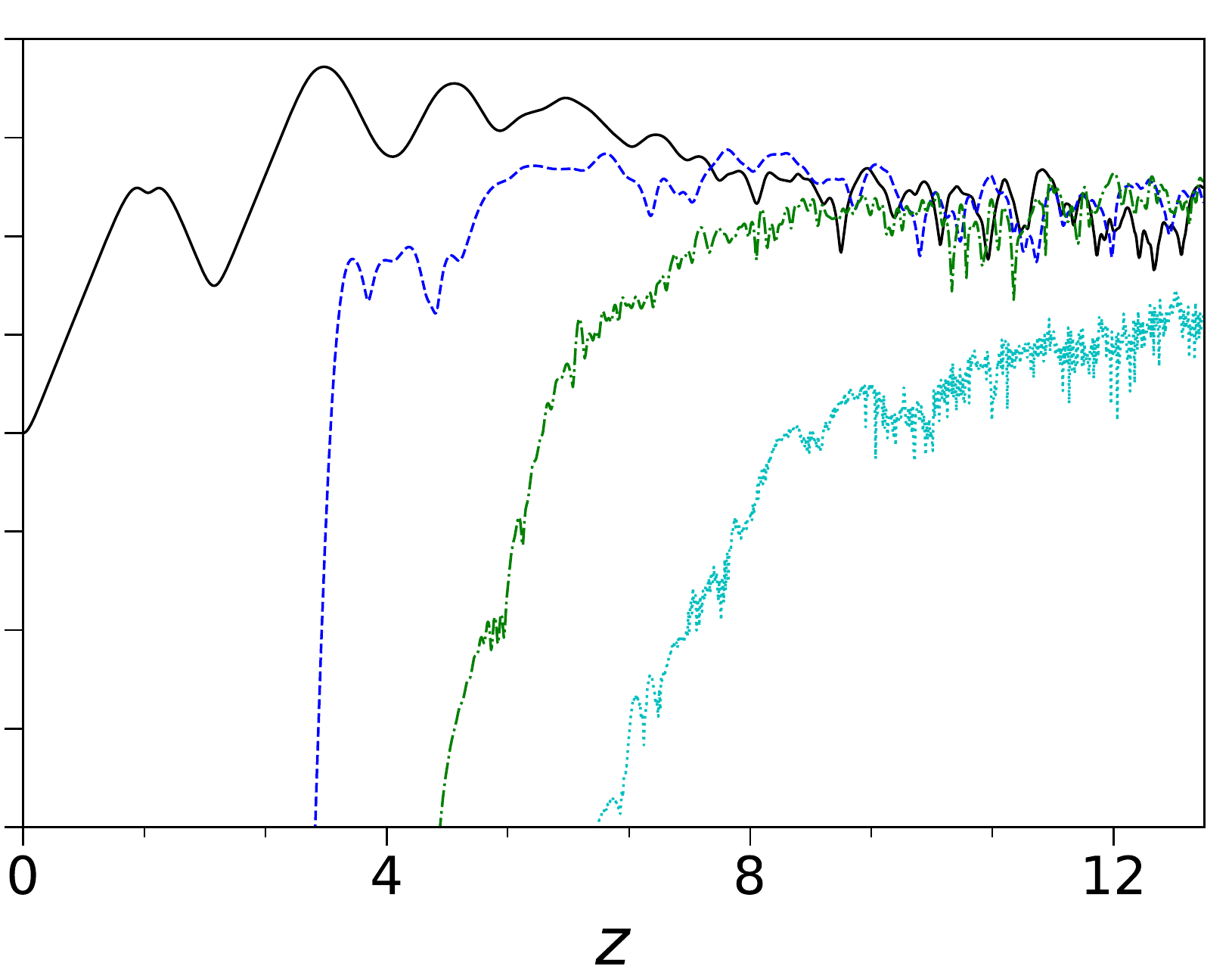} &
      \includegraphics*[scale=\figscalei]{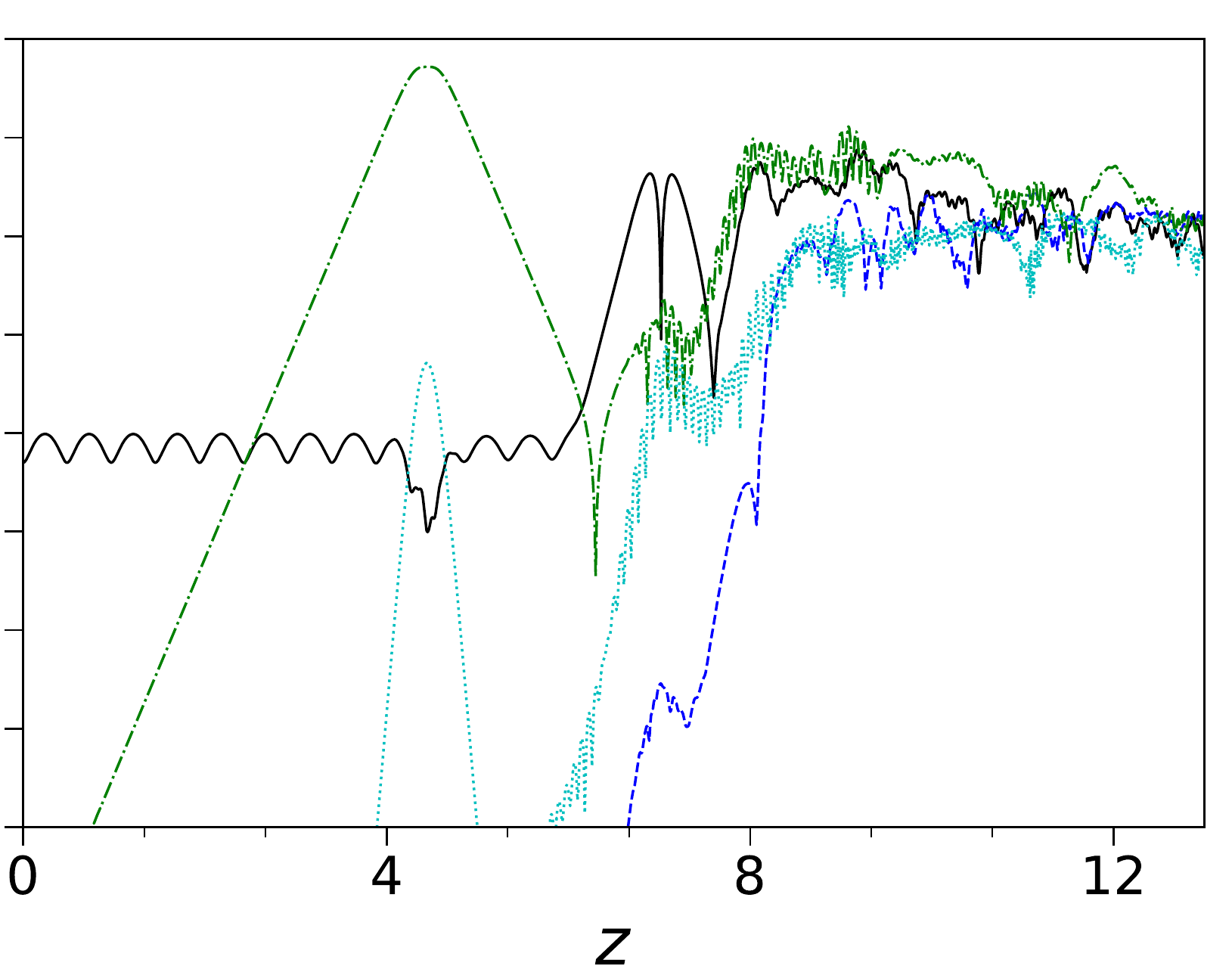}
    \end{array}$
  \end{center}
  \caption{(Color online)     \label{fig:moments}
    The development of the small-scale flavor structures in the
    neutrino gases with the neutrino self-coupling strengths $\mu=10$
    (left panels), $25$ (middle panels) and
    $50$ (right panels). All the calculations assume the inverted
    neutrino mass hierarchy and the antineutrino-to-neutrino density
    ratio $\alpha=0.6$. The top panels show $\avg{P_3}$ as functions
    of $x$ and $z$. The lower two rows show the magnitudes of
    the Fourier moments of the neutrino polarization vector
    $|\avg{\pol^{(m)}}|$ as functions of the
    Fourier index $m$ for a few values of $z$ (middle row), and as
    functions of $z$ for a few moments (bottom row). Also shown in the
    middle panels are the maximum exponential growth rates $\kappa$ as
    functions of $m$ (dashed curves with the scales on the right sides
    of the panels).
  }
\end{figure*}

\begin{figure*}[htb]
  \begin{center}
    $\begin{array}{@{}l@{\hspace{0.01in}}l@{\hspace{0.01in}}l@{}}
      \includegraphics*[scale=\figscalei]{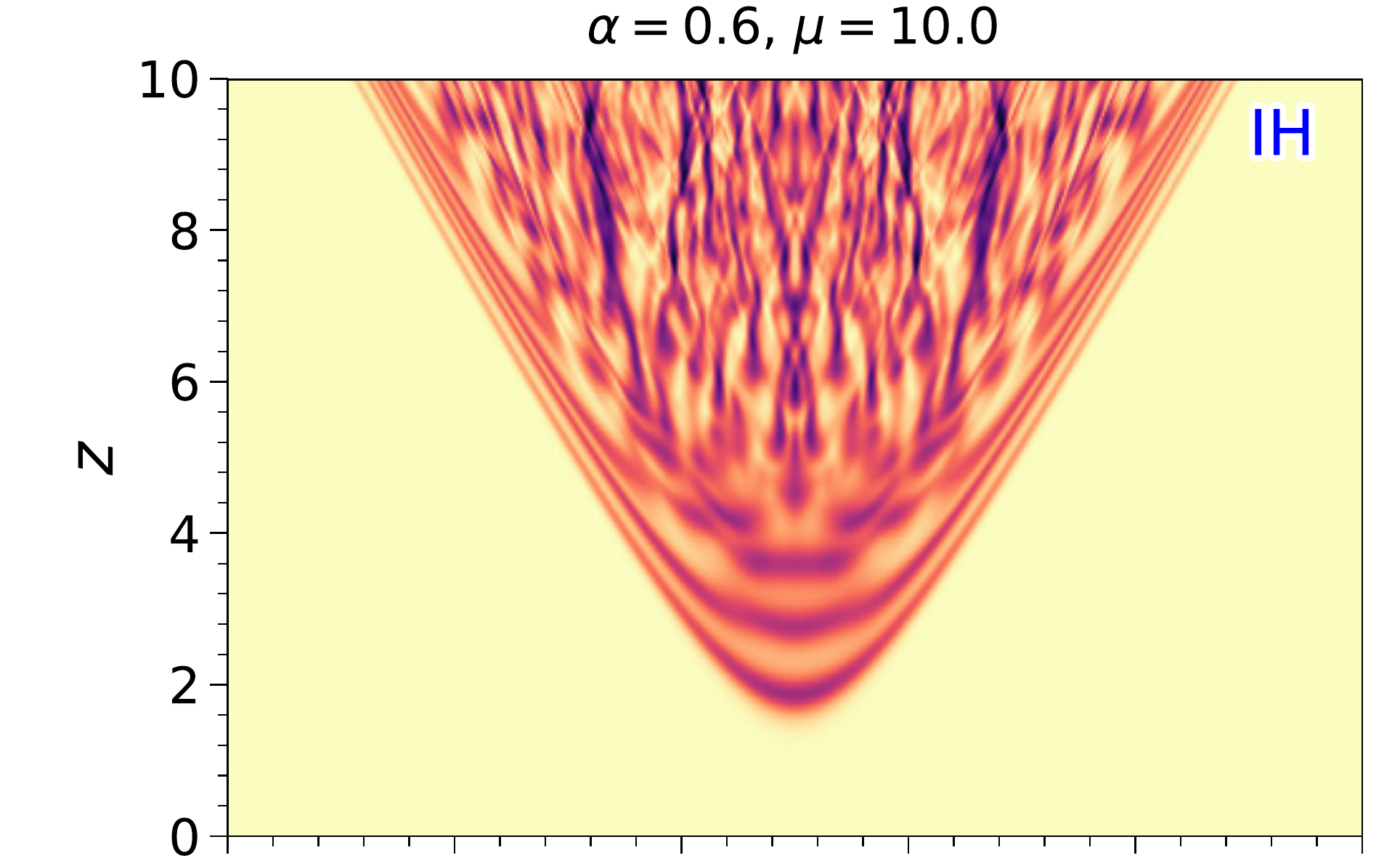} &
      \includegraphics*[scale=\figscalei]{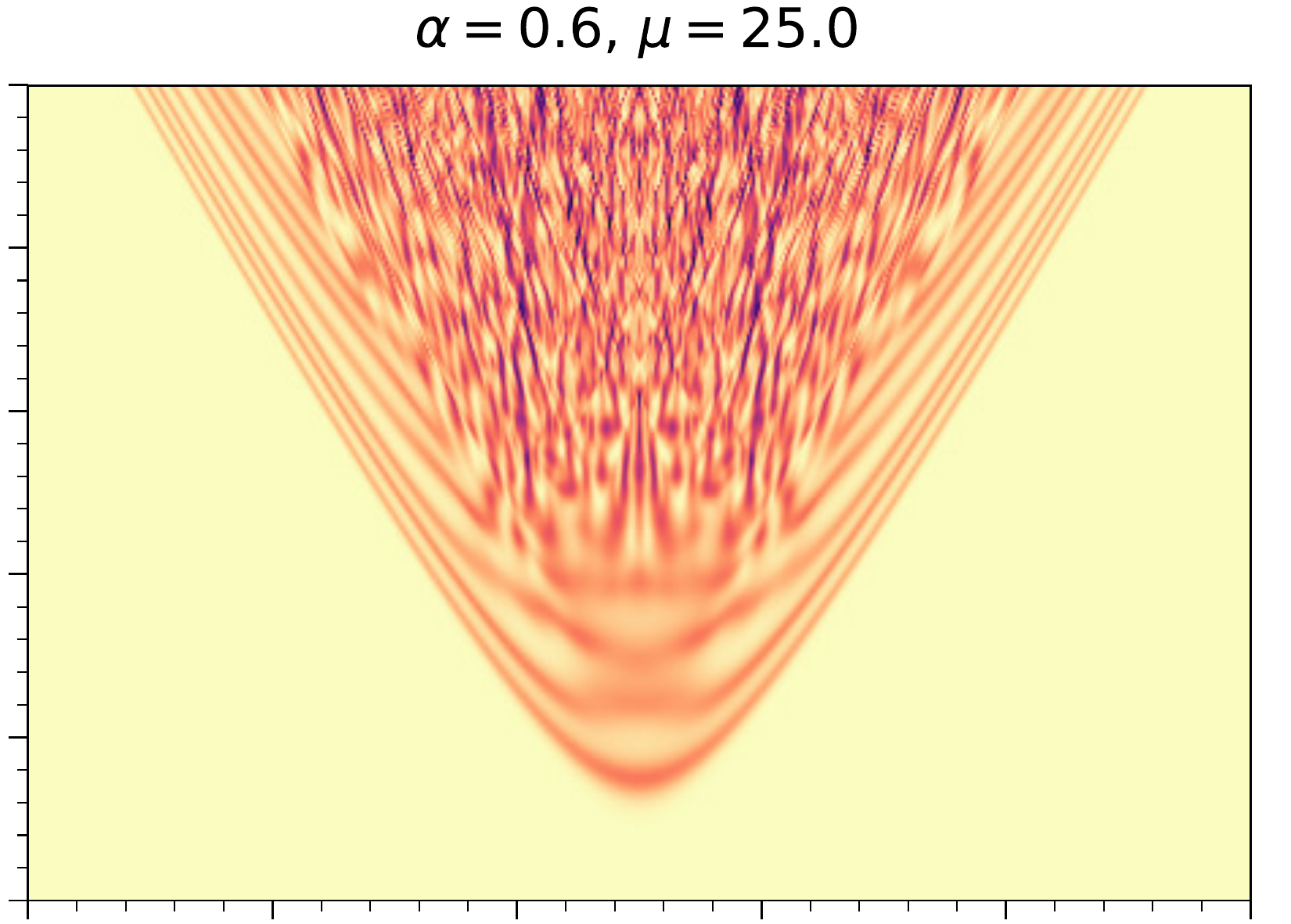} &
      \includegraphics*[scale=\figscalei]{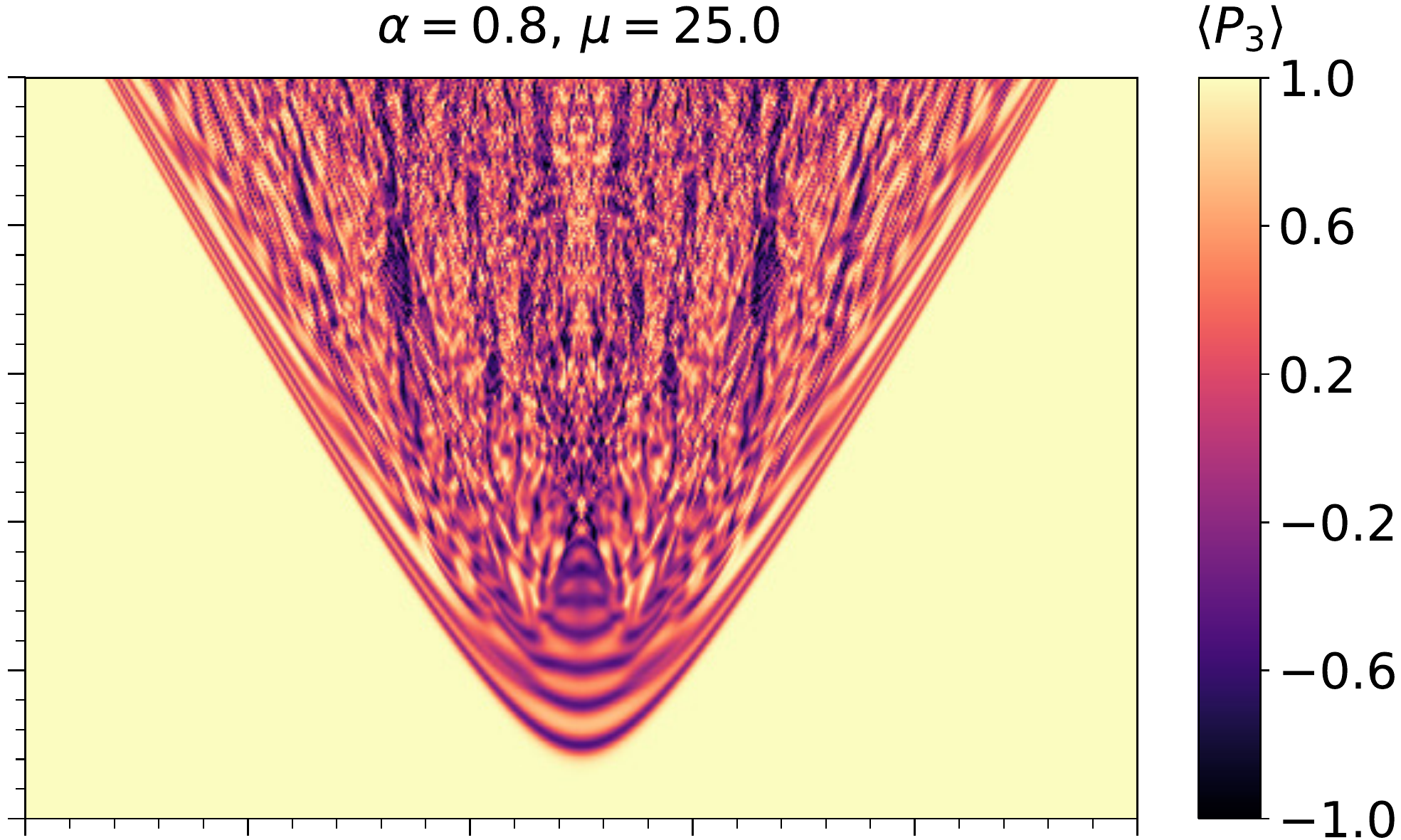} \\
      \includegraphics*[scale=\figscalei]{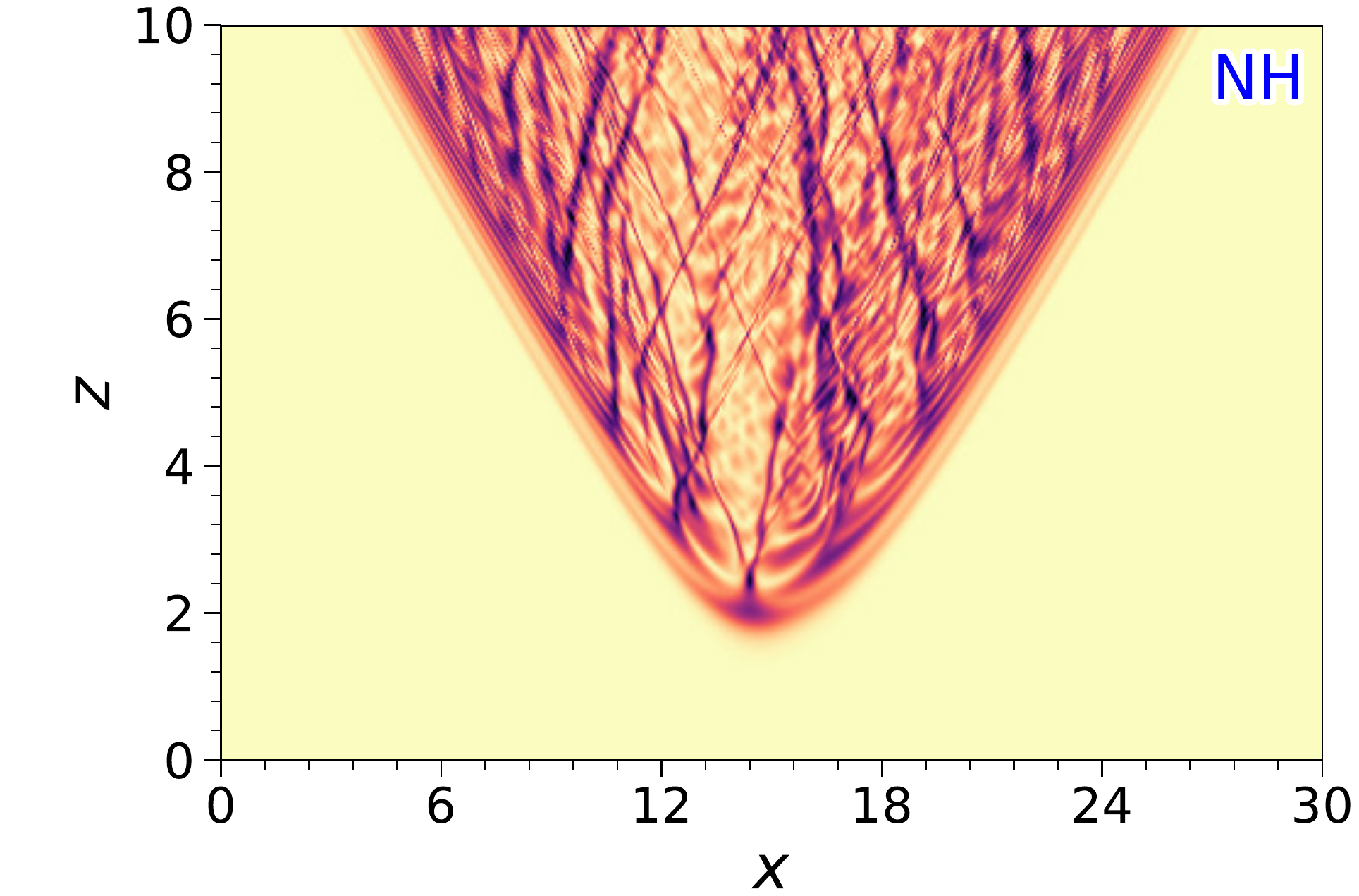} &
      \includegraphics*[scale=\figscalei]{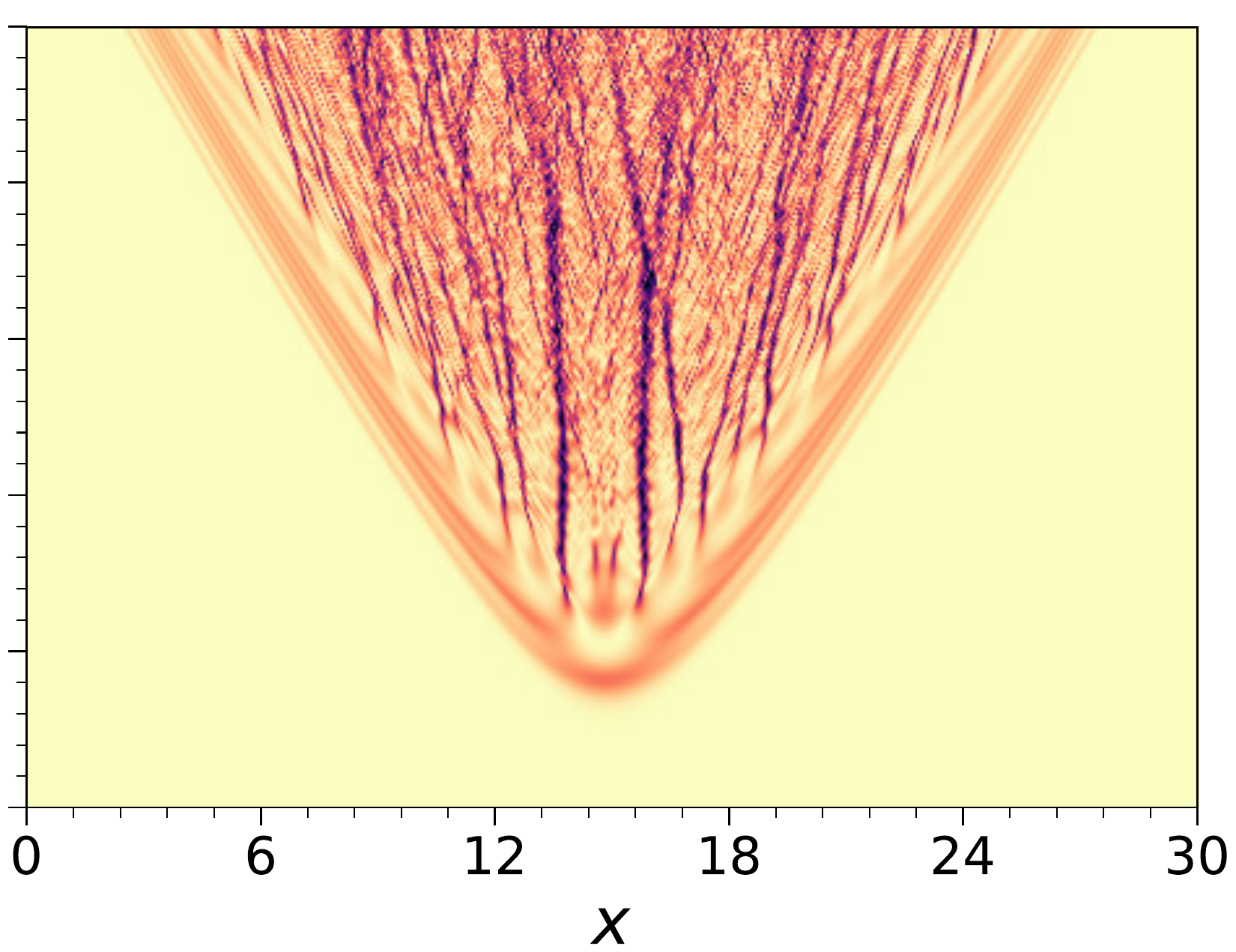} &
      \includegraphics*[scale=\figscalei]{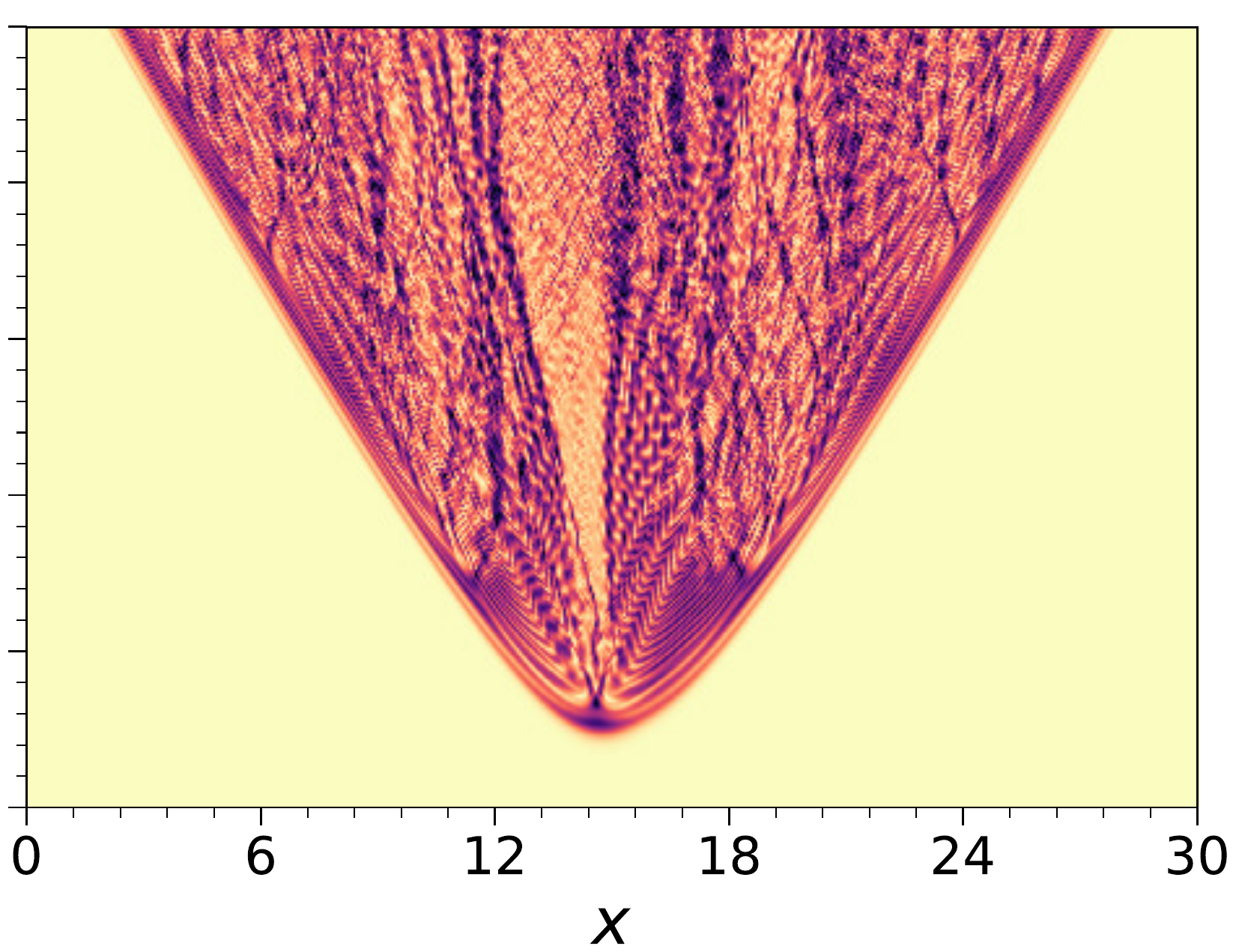} \\
      \includegraphics*[scale=\figscalei]{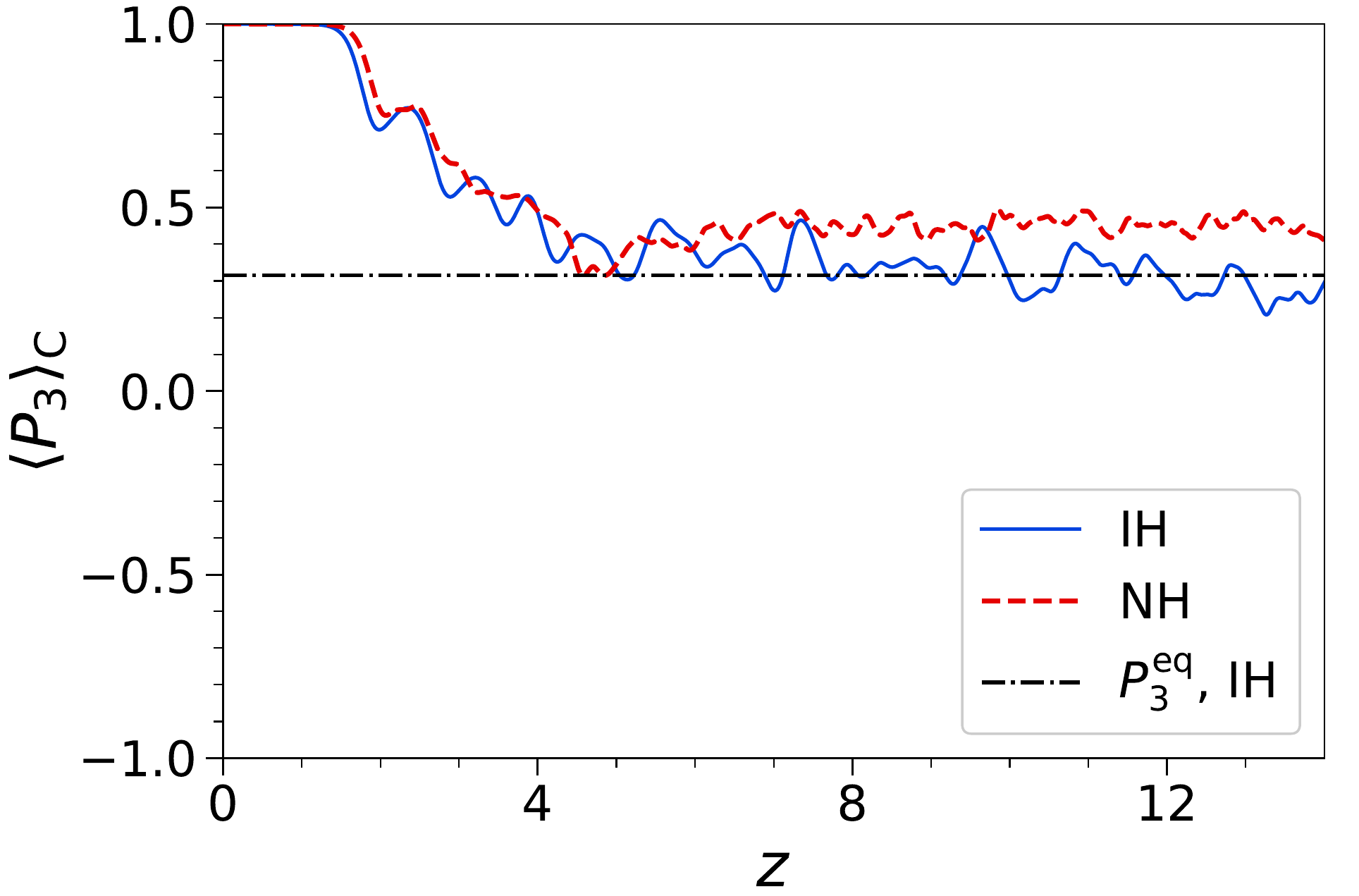} &
      \includegraphics*[scale=\figscalei]{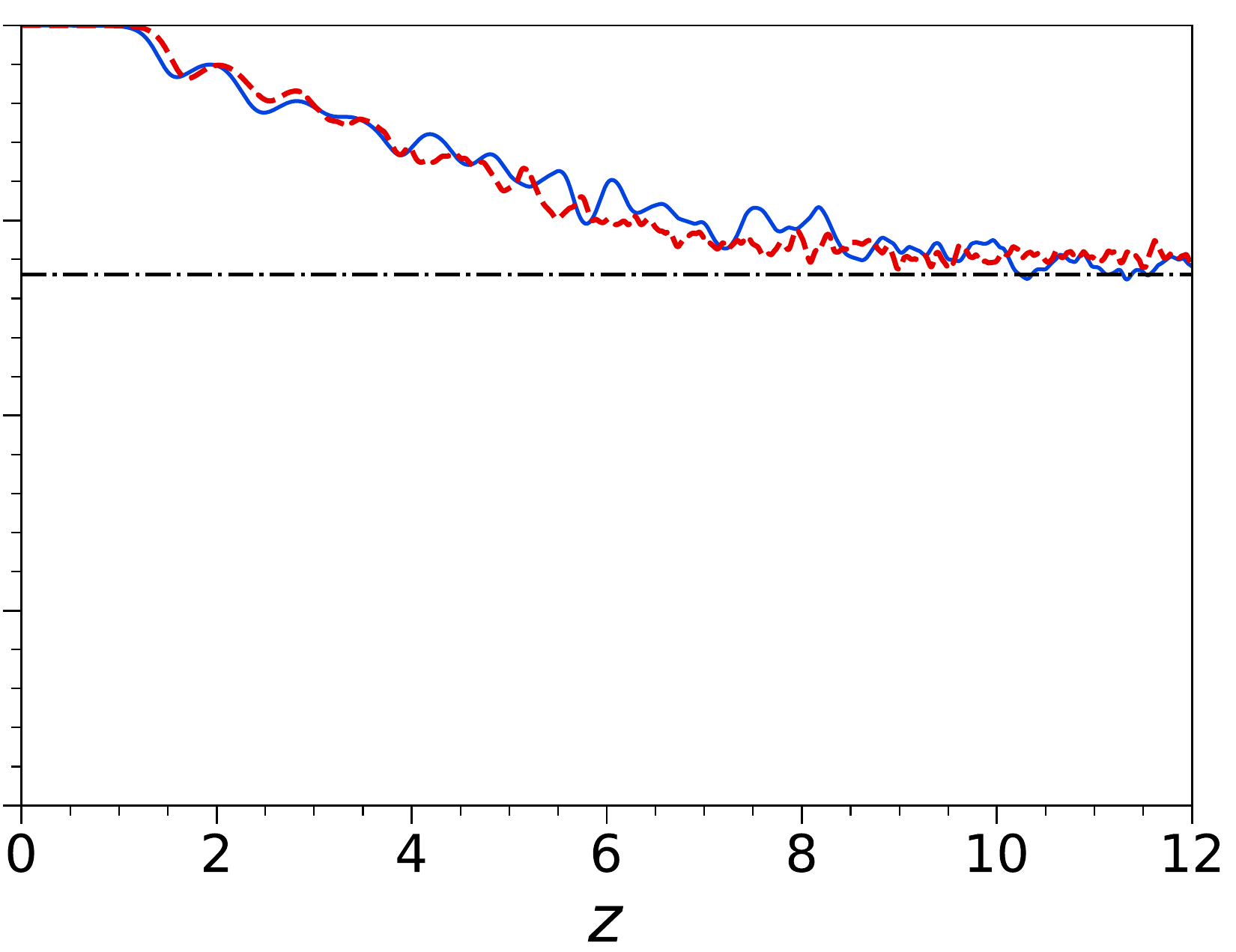} &
      \includegraphics*[scale=\figscalei]{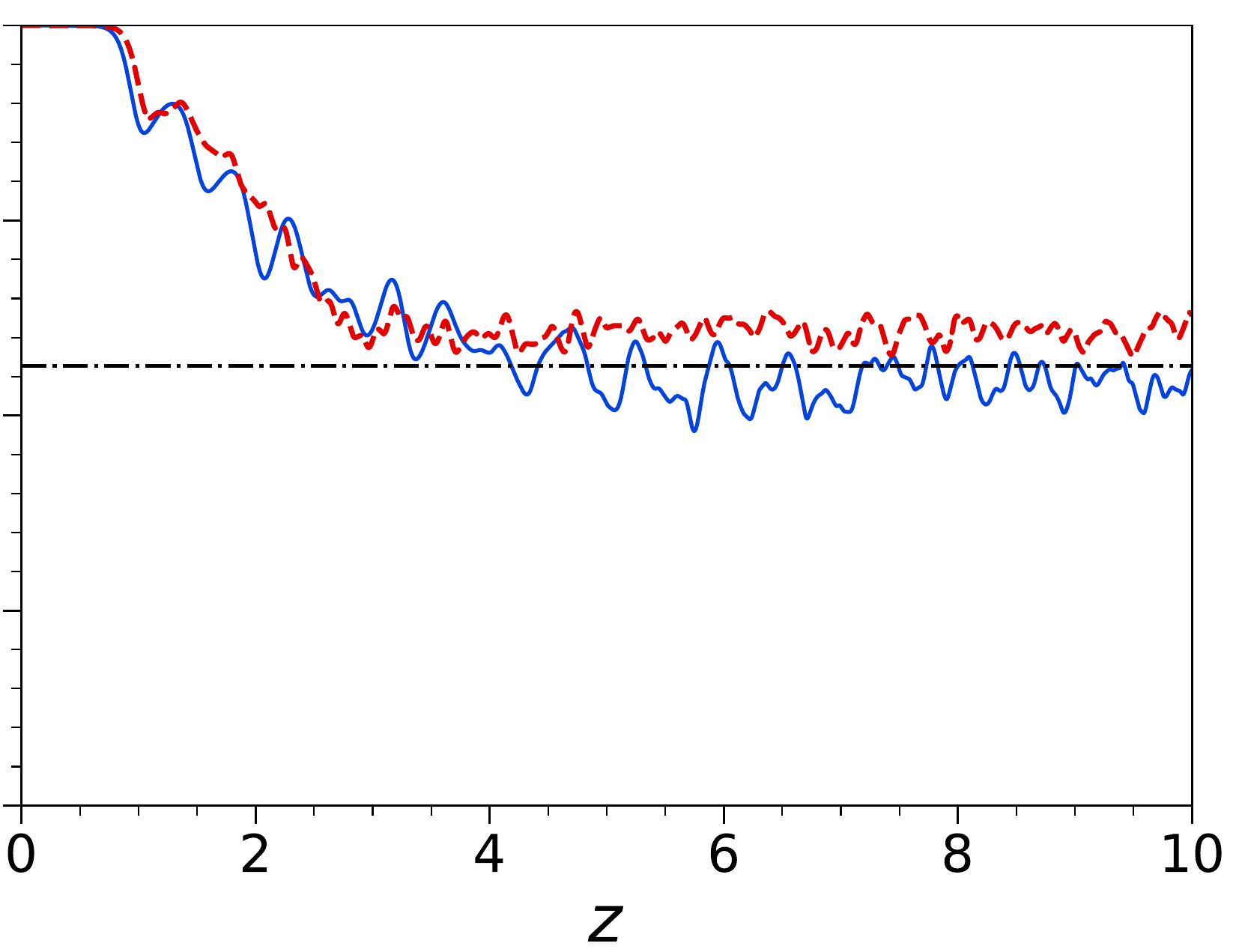}
    \end{array}$
  \end{center}
  \caption{(Color online)     \label{fig:P3-gauss}
    Similar to Fig.~\ref{fig:P3-sine} but for highly localized initial
    perturbations (see text) and different values of $\alpha$ and
    $\mu$. In the bottom row, $\avg{P_3}_\text{C}$ is averaged over
    the central segment of the box with a width $10$. Also shown are $P_3^\eq$
    in the corresponding IH calculations with sinusoidal initial
    perturbations.
  }
\end{figure*}

We first consider the neutrino gas with a sinusoidal initial
perturbation:
\begin{align}
  \epsilon_\pm(x) &= \epsilon_0^\pm
  + 2\sum_{m>0} \epsilon_m^\pm \sin(k_m x),
  \label{eq:sinIC}
\end{align}
where $\epsilon_m^\pm$ are constants. The results from this idealized
initial condition demonstrates how the flavor evolution in the
two-beam neutrino line model depends on the neutrino mass hierarchy,
the neutrino-antineutrino
asymmetry, and the neutrino number density.

In Fig.~\ref{fig:P3-sine} we show $\avg{P_3(x,z)}$ with $\mu=25$ and
three values of $\alpha$ for both $\eta=-1$ (IH, top panels) and
$\eta=+1$ (NH, middle panels), where $\avg{\cdots}$ represents the
average over the emission directions of the neutrino fluxes.
In the bottom panels of the figure, we show the evolution of $P_3$
when averaged over both the neutrino emission directions and the $x$
axis.

In the IH cases, we set $\epsilon_m^\pm=0$ for all $m$ except
\begin{align}
  \epsilon_0^\pm=2\times10^{-3}
  \quad\text{and}\quad
  \epsilon_1^\pm=-10^{-4}.
  \label{eq:eps-IH}
\end{align}
Although the left- and right-going neutrino beams have the same initial
conditions at any emission point on the $x$ axis, this left-right
symmetry is lost at $z>0$ when the $x$-translation symmetry is
spontaneously broken. When averaged over the left- and right-going beams,
however, $\avg{P_3(x,z)}$ demonstrates a residue mirror symmetry about the
middle line of the box because of the left-right symmetry in the
initial condition.
The overall flavor evolution patterns displayed in the top panels of
Fig.~\ref{fig:P3-sine} with different values of
$\alpha$ are very similar. Initially, the neutrino gas behaves like a flavor
pendulum: the neutrino polarization vectors remain in their initial
states for a long time before quickly swinging towards the opposite
flavor states and coming back up again. The approximate
$x$-translation symmetry is preserved during this pendulum-like flavor
evolution until the $m=1$ mode becomes significant.
After that, the large-scale flavor
structures along the $x$ direction begin to break down into
small-scale structures as $z$ increases. The overall flavor conversion
at large $z$ increases with $\alpha$ for a fixed value of $\mu$.

The initial conditions in the NH cases shown in the
middle panels of Fig.~\ref{fig:P3-sine} are the same as in the IH cases
except
\begin{align}
  \epsilon_1^-=2\epsilon_1^+=-2\times10^{-4}.
\end{align}
Because the homogeneous mode is perturbed symmetrically
(i.e., $\epsilon_0^+=\epsilon_0^-$) and
because the flavor pendulum is stable with $\eta=+1$, the neutrino gas
does not experience significant flavor conversion until the $m=1$ mode
becomes significant. Similar to the IH cases, the
large-scale flavor structures also break down into small-scale ones in
the NH cases as $z$
increases. But compared to their IH counterparts, $\avg{P_3(x,z)}$ with
the NH develop more prominent stream-like structures that are
localized in the $x$ direction and extended along the $z$
direction. These stream-like flavor structures result in the
``streamlines of the neutrino flux'' observed in
Ref.~\cite{Mirizzi:2015fva}.

The results shown in the bottom panels of Fig.~\ref{fig:P3-sine}
suggest that, although the
initial flavor evolution of the neutrino gas can be sensitive to the
neutrino mass hierarchy and the initial condition at $z=0$, the
overall flavor conversion at large $z$ appears to settle down on an
equilibrium value which is almost independent of these
parameters. In practice, we define $P^\eq_3$ and $\Pb^\eq_3$ to be the
mean values
of $\avg{P_3^{(0)}(z)}$ and $\avg{\Pb_3^{(0)}(z)}$ in the last
distance unit of $z$ after they are saturated.

To demonstrate the dependence of $P^\eq_3$ and $\Pb^\eq_3$ on the
neutrino-antineutrino asymmetry and the neutrino density,
we plot $P^\eq_3$ and $\Pb^\eq_3$ in Fig.~\ref{fig:P3eq} in terms of
$\alpha$ (left panel) and $\mu$ (right panel), respectively, both with
$\eta=-1$ and the initial condition defined in
Eq.~\eqref{eq:eps-IH}. From Fig.~\ref{fig:P3eq} one can see that the
antineutrinos in the two-beam line model become almost depolarized in
flavor (i.e., $\Pb^\eq_3\approx 0$) at large $z$. The depolarization
is generally more complete with a larger value of $\mu$ or
neutrino density. Because of the
conserved electron lepton number $\mathcal{L}$ [Eq.~\eqref{eq:L}],
the final mean flavor conversion of the neutrinos is related to
that of the antineutrinos through
\begin{align}
  P^\eq_3  = 1 - \alpha (1 - \Pb^\eq_3).
  \label{eq:P3eq-cons}
\end{align}
As shown in Fig.~\ref{fig:P3eq}, the values of $P^\eq_3$ in our
calculations indeed demonstrate a nearly linear dependence on $\alpha$
but little dependence on $\mu$.

Although the overall flavor conversion of the neutrino gas in the
two-beam line
model is largely independent of the neutrino self-coupling strength
$\mu$, a larger number of $x$ bins is needed to achieve the same
accuracy as $\mu$ increases (see the changing symbols in the right
panel of Fig.~\ref{fig:P3eq}). This is because more significant
small-scale flavor structures are generated at smaller $z$ as $\mu$
increases. To demonstrate this interesting feature, we show the flavor
evolution of the neutrino gas with the same neutrino-antineutrino
asymmetry ($\alpha=0.6$) but three
different values of $\mu$ in Fig.~\ref{fig:moments}. All three
calculations assume the inverted neutrino mass hierarchy. The initial
flavor perturbations of the neutrino gas in these calculations are the
same as what is defined in Eq.~\eqref{eq:eps-IH} except with
\begin{align}
  \epsilon_0^\pm=10^{-3},\,
  \epsilon_1^\pm=5\times10^{-5}
    \,\text{and}\,
    \epsilon_{200}^\pm=5\times10^{-10}
  \label{eq:eps-IH}
\end{align}
for the case with $\mu=50$.

Comparing the first two panels of Fig.~\ref{fig:moments} in the first
row, one can clearly see that the neutrino gas with a larger value of
$\mu$ or a higher
neutrino density develops finer flavor structures than the one with
a lower density. This fact becomes obvious when one compares the magnitudes
the Fourier moments of the neutrino polarization vector
$|\avg{\pol^{(m)}}|$ in these two cases, as shown in the middle panels
of Fig.~\ref{fig:moments}. It appears that $|\avg{\pol^{(m)}}|$ has a
semi-exponential dependence on the Fourier index $m$ for a wide range
of $m$ and $z$:
\begin{align}
|\avg{\pol^{(m)}}| \propto \exp[-\beta(z)\, k_m],
\end{align}
where the exponent $\beta(z)$ decreases with increasing $z$. The
comparison between the first two panels in the second row of
Fig.~\ref{fig:moments} shows that $\beta$ is smaller at the same $z$ for
the neutrino gas with a larger density which indicates more prominent
small-scale flavor structures.

The semi-exponential form of $|\avg{\pol^{(m)}}|$ in the nonlinear
regime is intriguing and cannot be explained by the linearized flavor
stability analysis.
In the middle panels of Fig.~\ref{fig:moments} we also plot the
largest exponential growth rate $\kappa(m)$ (dashed curves) predicted by the
linearized flavor stability analysis. In the first two panels of this row,
one observes that the high moments rise faster than the low moments in
order to maintain the semi-exponential power spectrum, while
$\kappa(m)$ decreases with increasing $m$. In fact, the magnitudes of
the Fourier moments with $m\gtrsim 200$ grow with $z$
for $\mu=10$ even though they are predicted to be stable in the linear
regime (left panel).
These results indicate that the excitation of the high moments in
these calculations is not due to
their own instabilities but because of the ``power diffusion'' from the low
moments. This diffusion phenomenon is
reminiscent of the development of the turbulence in an initially
laminarly flowing fluid  \cite{Mirizzi:2015fva} and the kinematic
decoherence in an initially isotropic neutrino gas \cite{Raffelt:2007yz}.

In the bottom panels of
Fig.~\ref{fig:moments} we also show the evolution of the strengths of a
few Fourier moments over $z$. These panels show that different Fourier
moments do not reach equilibrium at the same time. The low
moments are saturated first, and the high moments later. This implies
that small-scale flavor structures may still develop even after the
overall flavor conversion of the neutrino gas has reached its
equilibrium value.

The calculations shown in the last column of Fig.~\ref{fig:moments} is special
because the flavor pendulum with $\alpha=0.6$ and
$\mu=50$ is in the sleeping-top regime, and the Fourier modes with
$m\lesssim 50$ are stable (see Fig.~\ref{fig:kappa}).
Therefore, we intentionally perturbed the Fourier modes with
$m=\pm200$ which exhibit the exponential growth in the linear regime
(bottom panel). After these moments are excited, they couple the low
moments to the high moments and, as a result, $|\avg{\pol^{(m)}}|$ has
semi-periodic dependence on $m$ initially (middle panel). At large
$z$, however, $|\avg{\pol^{(m)}}|$ in this case has a much weaker
dependence on $m$ than those shown in the left two panels.

\subsection{Localized perturbations} \label{subsec:gaussIC}

\newcommand{\epsg}{\epsilon_\text{g}}
The sinusoidal initial perturbations discussed in the previous subsection
are large-scale perturbations. Next we consider localized initial
perturbations of the form
\begin{align}
  \epsilon_\pm(x) = \frac{\epsg^\pm}{\sqrt{2\pi\sigma^2}}\,
  \exp\left[-\frac{(x-x_0)^2}{2\sigma^2}\right],
\end{align}
where $\epsg^\pm$, $x_0$, and $\sigma$ are constants. In
Fig.~\ref{fig:P3-gauss} we show the flavor evolution of the neutrino
gases with various values of $\alpha$ and $\mu$ for both the inverted
(IH, top panels) and normal (NH, middle panels) neutrino mass hierarchies.
For the initial perturbations, we take
\begin{align}
  \epsg^\pm &= 2\times10^{-3}, &
  x_0 &= L/2, &
  \sigma^2 &= 0.2
\end{align}
for the IH cases, and the same initial
perturbations except with
\begin{align}
  \epsg^- = 2 \epsg^+ = 4\times 10^{-3}
\end{align}
for the NH cases.

The upper two rows of Fig.~\ref{fig:P3-gauss} show that, for an
initial perturbation localized around $x_0$, the flavor conversion
first appears near $x_0$ and spreads out to the left and right. At
large $z$ the edges of the ``envelope'' of the flavor
conversion region are parallel to the propagation directions of the
two velocity modes of the neutrinos. The flavor conversion is
coherent in space near these edges, but small-scale flavor
structures develop deep inside the envelope.
The flavor development inside the envelope shown in
Fig.~\ref{fig:P3-gauss} is qualitatively the same as that shown in
Figs.~\ref{fig:P3-sine} and \ref{fig:moments}. A larger neutrino
self-coupling strength or
density induces finer flavor structures, and a smaller
neutrino-antineutrino asymmetry produces a larger overall flavor
conversion.

We intentionally limit our calculations to $z < L/2$ to avoid the
unphysical consequences because of the artificial, periodic
boundary condition. As a result,
the flavor evolution of the neutrino gases with localized
perturbations do not reach an equilibrium in the whole periodic box because
unconverted neutrinos continue to stream into the envelope. Instead of
averaging over the whole box, we average $\avg{P_3}$ over the central
region of the envelope and define
\begin{align}
  \avg{P_3(z)}_\text{C} = \frac{1}{\Delta L}\int_{x_0-\Delta L/2}^{x_0+\Delta L/2}
  \avg{P_3(x,z)}\,\dd x
\end{align}
with $\Delta L=L/3$.
In the bottom panels of Fig.~\ref{fig:P3-gauss} we compare the values of
$\avg{P_3(z)}_\text{C}$ in both the IH and NH cases and $P_e^\eq$
obtained from the IH calculations with sinusoidal
perturbations. Our calculations suggest that the overall flavor
conversions in the central region of the envelope are approximately
the same as those in the neutrino gases with the sinusoidal initial
perturbations. However, we also notice the NH cases tend to develop
flavor structures of the shapes of ``streams'' and ``domains'' as
observed in the previous study \cite{Mirizzi:2015fva}. Such regions
can have interesting physical consequences if developed in a real
astrophysical scenario.

\section{Conclusions} \label{sec:conclusion}

We have developed a numerical code to solve the 2D
($x$-$z$), two-beam
neutrino line model. We studied the flavor development of the neutrino
gases with small sinusoidal and localized perturbations initially (at $z=0$).

The neutrino gases with small sinusoidal perturbations behave
like a flavor pendulum initially which is coherent in the $x$
direction (along which
the gas is almost homogeneous initially) when the neutrino density is below a
critical value.  This coherent flavor evolution breaks down into small
flavor structures as the neutrinos propagate in the $z$ direction.
The Fourier analysis shows that the magnitudes of the flavor
structures have a semi-exponential dependence on the wave number
(in the $x$ direction) and change with $z$. The
overall flavor conversions in the neutrino gases (averaged over the $x$
axis) eventually achieve constant equilibrium values at large $z$.
In our calculations where
there are fewer antineutrinos than neutrinos, the antineutrinos almost
reach equipartition among different flavors, and the overall flavor conversions
of the neutrinos are approximately determined by the
neutrino-antineutrino asymmetries  according the conservation law of
the electron lepton number. This general behavior is largely independent of
the neutrino densities or the neutrino mass hierarchy.
However, the neutrino gas tend to develop prominent stream- and
domain-like flavor structures when the neutrino mass hierarchy is
normal. Also the increase of the neutrino density
causes the development of more prominent fine flavor structures at
smaller $z$ which make the problem more difficult to solve.

The flavor evolution in a neutrino gas with an initially localized
perturbation starts from the $x$ coordinate where the perturbation is
located and expands afterwards. The flavor development inside the
region where the oscillations occur is
very similar to that of a neutrino gas with an initial sinusoidal
perturbation.

The qualitative results obtained in our numerical survey, such as the
semi-exponential power spectrum of the flavor conversion and the final
equilibrium conversion probabilities,
are very intriguing.
However, it remains to be seen whether these results will survive in the
more sophisticated models (e.g., the line model with multiple neutrino
beams \cite{Mirizzi:2015fva} or the ring model with position dependent
neutrino densities \cite{Mirizzi:2015hwa}). If they do, then these
results suggest that some simple analytic
understanding and statistical treatment may be possible for the flavor
oscillations in a multi-dimensional neutrino gas.
Such a treatment would be
extremely useful in computing the neutrino signals and nucleosynthesis
in astrophysical scenarios such as core-collapse supernovae and
neutron star mergers.

\section*{Acknowledgments}
We thank J.~Carlson, L.~Ma, E.~Putney, S.~Shalgar, and C.~Yi for
useful discussions.
This work is supported by the US DOE EPSCoR grant DE-SC0008142
(S.~A.\ and H.~D.) and the NP
grant DE-SC0017803 at UNM (J.~D.~M.\ and H.~D.).

\bibliography{lineModel}

\begin{thebibliography}{36}%
\makeatletter
\providecommand \@ifxundefined [1]{%
 \@ifx{#1\undefined}
}%
\providecommand \@ifnum [1]{%
 \ifnum #1\expandafter \@firstoftwo
 \else \expandafter \@secondoftwo
 \fi
}%
\providecommand \@ifx [1]{%
 \ifx #1\expandafter \@firstoftwo
 \else \expandafter \@secondoftwo
 \fi
}%
\providecommand \natexlab [1]{#1}%
\providecommand \enquote  [1]{``#1''}%
\providecommand \bibnamefont  [1]{#1}%
\providecommand \bibfnamefont [1]{#1}%
\providecommand \citenamefont [1]{#1}%
\providecommand \href@noop [0]{\@secondoftwo}%
\providecommand \href [0]{\begingroup \@sanitize@url \@href}%
\providecommand \@href[1]{\@@startlink{#1}\@@href}%
\providecommand \@@href[1]{\endgroup#1\@@endlink}%
\providecommand \@sanitize@url [0]{\catcode `\\12\catcode `\$12\catcode
  `\&12\catcode `\#12\catcode `\^12\catcode `\_12\catcode `\%12\relax}%
\providecommand \@@startlink[1]{}%
\providecommand \@@endlink[0]{}%
\providecommand \url  [0]{\begingroup\@sanitize@url \@url }%
\providecommand \@url [1]{\endgroup\@href {#1}{\urlprefix }}%
\providecommand \urlprefix  [0]{URL }%
\providecommand \Eprint [0]{\href }%
\providecommand \doibase [0]{http://dx.doi.org/}%
\providecommand \selectlanguage [0]{\@gobble}%
\providecommand \bibinfo  [0]{\@secondoftwo}%
\providecommand \bibfield  [0]{\@secondoftwo}%
\providecommand \translation [1]{[#1]}%
\providecommand \BibitemOpen [0]{}%
\providecommand \bibitemStop [0]{}%
\providecommand \bibitemNoStop [0]{.\EOS\space}%
\providecommand \EOS [0]{\spacefactor3000\relax}%
\providecommand \BibitemShut  [1]{\csname bibitem#1\endcsname}%
\let\auto@bib@innerbib\@empty
\bibitem [{\citenamefont {Tanabashi}\ \emph {et~al.}(2018)\citenamefont
  {Tanabashi} \emph {et~al.}}]{Tanabashi:2018oca}%
  \BibitemOpen
  \bibfield  {author} {\bibinfo {author} {\bibfnamefont {M.}~\bibnamefont
  {Tanabashi}} \emph {et~al.} (\bibinfo {collaboration} {Particle Data
  Group}),\ }\bibfield  {title} {\enquote {\bibinfo {title} {{Review of
  Particle Physics}},}\ }\href {\doibase 10.1103/PhysRevD.98.030001} {\bibfield
   {journal} {\bibinfo  {journal} {Phys. Rev.}\ }\textbf {\bibinfo {volume}
  {D98}},\ \bibinfo {pages} {030001} (\bibinfo {year} {2018})}\BibitemShut
  {NoStop}%
\bibitem [{\citenamefont {Wolfenstein}(1978)}]{Wolfenstein:1977ue}%
  \BibitemOpen
  \bibfield  {author} {\bibinfo {author} {\bibfnamefont {L.}~\bibnamefont
  {Wolfenstein}},\ }\bibfield  {title} {\enquote {\bibinfo {title} {{Neutrino
  Oscillations in Matter}},}\ }\href {\doibase 10.1103/PhysRevD.17.2369}
  {\bibfield  {journal} {\bibinfo  {journal} {Phys. Rev.}\ }\textbf {\bibinfo
  {volume} {D17}},\ \bibinfo {pages} {2369--2374} (\bibinfo {year} {1978})},\
  \bibinfo {note} {[,294(1977)]}\BibitemShut {NoStop}%
\bibitem [{\citenamefont {Mikheyev}\ and\ \citenamefont
  {Smirnov}(1985)}]{Mikheev:1986gs}%
  \BibitemOpen
  \bibfield  {author} {\bibinfo {author} {\bibfnamefont {S.~P.}\ \bibnamefont
  {Mikheyev}}\ and\ \bibinfo {author} {\bibfnamefont {A.}~\bibnamefont
  {Smirnov}},\ }\bibfield  {title} {\enquote {\bibinfo {title} {{Resonance
  Amplification of Oscillations in Matter and Spectroscopy of Solar
  Neutrinos}},}\ }\href@noop {} {\bibfield  {journal} {\bibinfo  {journal}
  {Sov. J. Nucl. Phys.}\ }\textbf {\bibinfo {volume} {42}},\ \bibinfo {pages}
  {913--917} (\bibinfo {year} {1985})},\ \bibinfo {note}
  {[,305(1986)]}\BibitemShut {NoStop}%
\bibitem [{\citenamefont {Fuller}\ \emph {et~al.}(1987)\citenamefont {Fuller},
  \citenamefont {Mayle}, \citenamefont {Wilson},\ and\ \citenamefont
  {Schramm}}]{Fuller:1987aa}%
  \BibitemOpen
  \bibfield  {author} {\bibinfo {author} {\bibfnamefont {George~M.}\
  \bibnamefont {Fuller}}, \bibinfo {author} {\bibfnamefont {Ron~W.}\
  \bibnamefont {Mayle}}, \bibinfo {author} {\bibfnamefont {James~R.}\
  \bibnamefont {Wilson}}, \ and\ \bibinfo {author} {\bibfnamefont {David~N.}\
  \bibnamefont {Schramm}},\ }\bibfield  {title} {\enquote {\bibinfo {title}
  {Resonant neutrino oscillations and stellar collapse},}\ }\href@noop {}
  {\bibfield  {journal} {\bibinfo  {journal} {Astrophys. J.}\ }\textbf
  {\bibinfo {volume} {322}},\ \bibinfo {pages} {795} (\bibinfo {year}
  {1987})}\BibitemShut {NoStop}%
\bibitem [{\citenamefont {Notzold}\ and\ \citenamefont
  {Raffelt}(1988)}]{Notzold:1987ik}%
  \BibitemOpen
  \bibfield  {author} {\bibinfo {author} {\bibfnamefont {Dirk}\ \bibnamefont
  {Notzold}}\ and\ \bibinfo {author} {\bibfnamefont {Georg}\ \bibnamefont
  {Raffelt}},\ }\bibfield  {title} {\enquote {\bibinfo {title} {{Neutrino
  Dispersion at Finite Temperature and Density}},}\ }\href {\doibase
  10.1016/0550-3213(88)90113-7} {\bibfield  {journal} {\bibinfo  {journal}
  {Nucl. Phys.}\ }\textbf {\bibinfo {volume} {B307}},\ \bibinfo {pages}
  {924--936} (\bibinfo {year} {1988})}\BibitemShut {NoStop}%
\bibitem [{\citenamefont {Pantaleone}(1992)}]{Pantaleone:1992xh}%
  \BibitemOpen
  \bibfield  {author} {\bibinfo {author} {\bibfnamefont {James~T.}\
  \bibnamefont {Pantaleone}},\ }\bibfield  {title} {\enquote {\bibinfo {title}
  {{Dirac neutrinos in dense matter}},}\ }\href {\doibase
  10.1103/PhysRevD.46.510} {\bibfield  {journal} {\bibinfo  {journal} {Phys.
  Rev.}\ }\textbf {\bibinfo {volume} {D46}},\ \bibinfo {pages} {510--523}
  (\bibinfo {year} {1992})}\BibitemShut {NoStop}%
\bibitem [{\citenamefont {Duan}\ \emph {et~al.}(2010)\citenamefont {Duan},
  \citenamefont {Fuller},\ and\ \citenamefont {Qian}}]{Duan:2010bg}%
  \BibitemOpen
  \bibfield  {author} {\bibinfo {author} {\bibfnamefont {Huaiyu}\ \bibnamefont
  {Duan}}, \bibinfo {author} {\bibfnamefont {George~M.}\ \bibnamefont
  {Fuller}}, \ and\ \bibinfo {author} {\bibfnamefont {Yong-Zhong}\ \bibnamefont
  {Qian}},\ }\bibfield  {title} {\enquote {\bibinfo {title} {{Collective
  Neutrino Oscillations}},}\ }\href {\doibase
  10.1146/annurev.nucl.012809.104524} {\bibfield  {journal} {\bibinfo
  {journal} {Ann. Rev. Nucl. Part. Sci.}\ }\textbf {\bibinfo {volume} {60}},\
  \bibinfo {pages} {569--594} (\bibinfo {year} {2010})},\ \Eprint
  {http://arxiv.org/abs/1001.2799} {arXiv:1001.2799 [hep-ph]} \BibitemShut
  {NoStop}%
\bibitem [{\citenamefont {Kostelecky}\ \emph {et~al.}(1993)\citenamefont
  {Kostelecky}, \citenamefont {Pantaleone},\ and\ \citenamefont
  {Samuel}}]{Kostelecky:1993yt}%
  \BibitemOpen
  \bibfield  {author} {\bibinfo {author} {\bibfnamefont {V.~Alan}\ \bibnamefont
  {Kostelecky}}, \bibinfo {author} {\bibfnamefont {James~T.}\ \bibnamefont
  {Pantaleone}}, \ and\ \bibinfo {author} {\bibfnamefont {Stuart}\ \bibnamefont
  {Samuel}},\ }\bibfield  {title} {\enquote {\bibinfo {title} {{Neutrino
  oscillation in the early universe}},}\ }\href {\doibase
  10.1016/0370-2693(93)90156-C} {\bibfield  {journal} {\bibinfo  {journal}
  {Phys. Lett.}\ }\textbf {\bibinfo {volume} {B315}},\ \bibinfo {pages}
  {46--50} (\bibinfo {year} {1993})}\BibitemShut {NoStop}%
\bibitem [{\citenamefont {Abazajian}\ \emph {et~al.}(2002)\citenamefont
  {Abazajian}, \citenamefont {Beacom},\ and\ \citenamefont
  {Bell}}]{Abazajian:2002qx}%
  \BibitemOpen
  \bibfield  {author} {\bibinfo {author} {\bibfnamefont {Kevork~N.}\
  \bibnamefont {Abazajian}}, \bibinfo {author} {\bibfnamefont {John~F.}\
  \bibnamefont {Beacom}}, \ and\ \bibinfo {author} {\bibfnamefont {Nicole~F.}\
  \bibnamefont {Bell}},\ }\bibfield  {title} {\enquote {\bibinfo {title}
  {{Stringent constraints on cosmological neutrino anti-neutrino asymmetries
  from synchronized flavor transformation}},}\ }\href {\doibase
  10.1103/PhysRevD.66.013008} {\bibfield  {journal} {\bibinfo  {journal} {Phys.
  Rev.}\ }\textbf {\bibinfo {volume} {D66}},\ \bibinfo {pages} {013008}
  (\bibinfo {year} {2002})},\ \Eprint {http://arxiv.org/abs/astro-ph/0203442}
  {arXiv:astro-ph/0203442 [astro-ph]} \BibitemShut {NoStop}%
\bibitem [{\citenamefont {Duan}\ \emph
  {et~al.}(2006{\natexlab{a}})\citenamefont {Duan}, \citenamefont {Fuller},
  \citenamefont {Carlson},\ and\ \citenamefont {Qian}}]{Duan:2006jv}%
  \BibitemOpen
  \bibfield  {author} {\bibinfo {author} {\bibfnamefont {Huaiyu}\ \bibnamefont
  {Duan}}, \bibinfo {author} {\bibfnamefont {George~M.}\ \bibnamefont
  {Fuller}}, \bibinfo {author} {\bibfnamefont {J.}~\bibnamefont {Carlson}}, \
  and\ \bibinfo {author} {\bibfnamefont {Yong-Zhong}\ \bibnamefont {Qian}},\
  }\bibfield  {title} {\enquote {\bibinfo {title} {{Coherent Development of
  Neutrino Flavor in the Supernova Environment}},}\ }\href {\doibase
  10.1103/PhysRevLett.97.241101} {\bibfield  {journal} {\bibinfo  {journal}
  {Phys. Rev. Lett.}\ }\textbf {\bibinfo {volume} {97}},\ \bibinfo {pages}
  {241101} (\bibinfo {year} {2006}{\natexlab{a}})},\ \Eprint
  {http://arxiv.org/abs/astro-ph/0608050} {arXiv:astro-ph/0608050 [astro-ph]}
  \BibitemShut {NoStop}%
\bibitem [{\citenamefont {Duan}\ \emph
  {et~al.}(2006{\natexlab{b}})\citenamefont {Duan}, \citenamefont {Fuller},
  \citenamefont {Carlson},\ and\ \citenamefont {Qian}}]{Duan:2006an}%
  \BibitemOpen
  \bibfield  {author} {\bibinfo {author} {\bibfnamefont {Huaiyu}\ \bibnamefont
  {Duan}}, \bibinfo {author} {\bibfnamefont {George~M.}\ \bibnamefont
  {Fuller}}, \bibinfo {author} {\bibfnamefont {J}~\bibnamefont {Carlson}}, \
  and\ \bibinfo {author} {\bibfnamefont {Yong-Zhong}\ \bibnamefont {Qian}},\
  }\bibfield  {title} {\enquote {\bibinfo {title} {{Simulation of Coherent
  Non-Linear Neutrino Flavor Transformation in the Supernova Environment. 1.
  Correlated Neutrino Trajectories}},}\ }\href {\doibase
  10.1103/PhysRevD.74.105014} {\bibfield  {journal} {\bibinfo  {journal} {Phys.
  Rev.}\ }\textbf {\bibinfo {volume} {D74}},\ \bibinfo {pages} {105014}
  (\bibinfo {year} {2006}{\natexlab{b}})},\ \Eprint
  {http://arxiv.org/abs/astro-ph/0606616} {arXiv:astro-ph/0606616 [astro-ph]}
  \BibitemShut {NoStop}%
\bibitem [{\citenamefont {Hannestad}\ \emph {et~al.}(2006)\citenamefont
  {Hannestad}, \citenamefont {Raffelt}, \citenamefont {Sigl},\ and\
  \citenamefont {Wong}}]{Hannestad:2006nj}%
  \BibitemOpen
  \bibfield  {author} {\bibinfo {author} {\bibfnamefont {Steen}\ \bibnamefont
  {Hannestad}}, \bibinfo {author} {\bibfnamefont {Georg~G.}\ \bibnamefont
  {Raffelt}}, \bibinfo {author} {\bibfnamefont {Gunter}\ \bibnamefont {Sigl}},
  \ and\ \bibinfo {author} {\bibfnamefont {Yvonne Y.~Y.}\ \bibnamefont
  {Wong}},\ }\bibfield  {title} {\enquote {\bibinfo {title} {{Self-induced
  conversion in dense neutrino gases: Pendulum in flavour space}},}\ }\href
  {\doibase 10.1103/PhysRevD.74.105010, 10.1103/PhysRevD.76.029901} {\bibfield
  {journal} {\bibinfo  {journal} {Phys. Rev.}\ }\textbf {\bibinfo {volume}
  {D74}},\ \bibinfo {pages} {105010} (\bibinfo {year} {2006})},\ \bibinfo
  {note} {[Erratum: Phys. Rev.D76,029901(2007)]},\ \Eprint
  {http://arxiv.org/abs/astro-ph/0608695} {arXiv:astro-ph/0608695 [astro-ph]}
  \BibitemShut {NoStop}%
\bibitem [{\citenamefont {Raffelt}\ and\ \citenamefont
  {Smirnov}(2007)}]{Raffelt:2007cb}%
  \BibitemOpen
  \bibfield  {author} {\bibinfo {author} {\bibfnamefont {Georg~G.}\
  \bibnamefont {Raffelt}}\ and\ \bibinfo {author} {\bibfnamefont
  {Alexei~{\relax Yu}.}\ \bibnamefont {Smirnov}},\ }\bibfield  {title}
  {\enquote {\bibinfo {title} {{Self-induced spectral splits in supernova
  neutrino fluxes}},}\ }\href {\doibase 10.1103/PhysRevD.76.081301,
  10.1103/PhysRevD.77.029903} {\bibfield  {journal} {\bibinfo  {journal} {Phys.
  Rev.}\ }\textbf {\bibinfo {volume} {D76}},\ \bibinfo {pages} {081301}
  (\bibinfo {year} {2007})},\ \bibinfo {note} {[Erratum: Phys.
  Rev.D77,029903(2008)]},\ \Eprint {http://arxiv.org/abs/0705.1830}
  {arXiv:0705.1830 [hep-ph]} \BibitemShut {NoStop}%
\bibitem [{\citenamefont {Raffelt}\ \emph {et~al.}(2013)\citenamefont
  {Raffelt}, \citenamefont {Sarikas},\ and\ \citenamefont
  {de~Sousa~Seixas}}]{Raffelt:2013rqa}%
  \BibitemOpen
  \bibfield  {author} {\bibinfo {author} {\bibfnamefont {Georg}\ \bibnamefont
  {Raffelt}}, \bibinfo {author} {\bibfnamefont {Srdjan}\ \bibnamefont
  {Sarikas}}, \ and\ \bibinfo {author} {\bibfnamefont {David}\ \bibnamefont
  {de~Sousa~Seixas}},\ }\bibfield  {title} {\enquote {\bibinfo {title} {{Axial
  Symmetry Breaking in Self-Induced Flavor Conversion of Supernova Neutrino
  Fluxes}},}\ }\href {\doibase 10.1103/PhysRevLett.113.239903,
  10.1103/PhysRevLett.111.091101} {\bibfield  {journal} {\bibinfo  {journal}
  {Phys. Rev. Lett.}\ }\textbf {\bibinfo {volume} {111}},\ \bibinfo {pages}
  {091101} (\bibinfo {year} {2013})},\ \bibinfo {note} {[Erratum: Phys. Rev.
  Lett.113,no.23,239903(2014)]},\ \Eprint {http://arxiv.org/abs/1305.7140}
  {arXiv:1305.7140 [hep-ph]} \BibitemShut {NoStop}%
\bibitem [{\citenamefont {Mirizzi}(2013)}]{Mirizzi:2013rla}%
  \BibitemOpen
  \bibfield  {author} {\bibinfo {author} {\bibfnamefont {Alessandro}\
  \bibnamefont {Mirizzi}},\ }\bibfield  {title} {\enquote {\bibinfo {title}
  {{Multi-azimuthal-angle effects in self-induced supernova neutrino flavor
  conversions without axial symmetry}},}\ }\href {\doibase
  10.1103/PhysRevD.88.073004} {\bibfield  {journal} {\bibinfo  {journal} {Phys.
  Rev.}\ }\textbf {\bibinfo {volume} {D88}},\ \bibinfo {pages} {073004}
  (\bibinfo {year} {2013})},\ \Eprint {http://arxiv.org/abs/1308.1402}
  {arXiv:1308.1402 [hep-ph]} \BibitemShut {NoStop}%
\bibitem [{\citenamefont {Duan}(2013)}]{Duan:2013kba}%
  \BibitemOpen
  \bibfield  {author} {\bibinfo {author} {\bibfnamefont {Huaiyu}\ \bibnamefont
  {Duan}},\ }\bibfield  {title} {\enquote {\bibinfo {title} {{Flavor
  Oscillation Modes In Dense Neutrino Media}},}\ }\href {\doibase
  10.1103/PhysRevD.88.125008} {\bibfield  {journal} {\bibinfo  {journal} {Phys.
  Rev.}\ }\textbf {\bibinfo {volume} {D88}},\ \bibinfo {pages} {125008}
  (\bibinfo {year} {2013})},\ \Eprint {http://arxiv.org/abs/1309.7377}
  {arXiv:1309.7377 [hep-ph]} \BibitemShut {NoStop}%
\bibitem [{\citenamefont {Mangano}\ \emph {et~al.}(2014)\citenamefont
  {Mangano}, \citenamefont {Mirizzi},\ and\ \citenamefont
  {Saviano}}]{Mangano:2014zda}%
  \BibitemOpen
  \bibfield  {author} {\bibinfo {author} {\bibfnamefont {Gianpiero}\
  \bibnamefont {Mangano}}, \bibinfo {author} {\bibfnamefont {Alessandro}\
  \bibnamefont {Mirizzi}}, \ and\ \bibinfo {author} {\bibfnamefont {Ninetta}\
  \bibnamefont {Saviano}},\ }\bibfield  {title} {\enquote {\bibinfo {title}
  {{Damping the neutrino flavor pendulum by breaking homogeneity}},}\ }\href
  {\doibase 10.1103/PhysRevD.89.073017} {\bibfield  {journal} {\bibinfo
  {journal} {Phys. Rev.}\ }\textbf {\bibinfo {volume} {D89}},\ \bibinfo {pages}
  {073017} (\bibinfo {year} {2014})},\ \Eprint {http://arxiv.org/abs/1403.1892}
  {arXiv:1403.1892 [hep-ph]} \BibitemShut {NoStop}%
\bibitem [{\citenamefont {Duan}\ and\ \citenamefont
  {Shalgar}(2015)}]{Duan:2014gfa}%
  \BibitemOpen
  \bibfield  {author} {\bibinfo {author} {\bibfnamefont {Huaiyu}\ \bibnamefont
  {Duan}}\ and\ \bibinfo {author} {\bibfnamefont {Shashank}\ \bibnamefont
  {Shalgar}},\ }\bibfield  {title} {\enquote {\bibinfo {title} {{Flavor
  instabilities in the neutrino line model}},}\ }\href {\doibase
  10.1016/j.physletb.2015.05.057} {\bibfield  {journal} {\bibinfo  {journal}
  {Phys. Lett.}\ }\textbf {\bibinfo {volume} {B747}},\ \bibinfo {pages}
  {139--143} (\bibinfo {year} {2015})},\ \Eprint
  {http://arxiv.org/abs/1412.7097} {arXiv:1412.7097 [hep-ph]} \BibitemShut
  {NoStop}%
\bibitem [{\citenamefont {Chakraborty}\ \emph
  {et~al.}(2016{\natexlab{a}})\citenamefont {Chakraborty}, \citenamefont
  {Hansen}, \citenamefont {Izaguirre},\ and\ \citenamefont
  {Raffelt}}]{Chakraborty:2015tfa}%
  \BibitemOpen
  \bibfield  {author} {\bibinfo {author} {\bibfnamefont {Sovan}\ \bibnamefont
  {Chakraborty}}, \bibinfo {author} {\bibfnamefont {Rasmus~Sloth}\ \bibnamefont
  {Hansen}}, \bibinfo {author} {\bibfnamefont {Ignacio}\ \bibnamefont
  {Izaguirre}}, \ and\ \bibinfo {author} {\bibfnamefont {Georg}\ \bibnamefont
  {Raffelt}},\ }\bibfield  {title} {\enquote {\bibinfo {title} {{Self-induced
  flavor conversion of supernova neutrinos on small scales}},}\ }\href
  {\doibase 10.1088/1475-7516/2016/01/028} {\bibfield  {journal} {\bibinfo
  {journal} {JCAP}\ }\textbf {\bibinfo {volume} {1601}},\ \bibinfo {pages}
  {028} (\bibinfo {year} {2016}{\natexlab{a}})},\ \Eprint
  {http://arxiv.org/abs/1507.07569} {arXiv:1507.07569 [hep-ph]} \BibitemShut
  {NoStop}%
\bibitem [{\citenamefont {Mirizzi}\ \emph {et~al.}(2015)\citenamefont
  {Mirizzi}, \citenamefont {Mangano},\ and\ \citenamefont
  {Saviano}}]{Mirizzi:2015fva}%
  \BibitemOpen
  \bibfield  {author} {\bibinfo {author} {\bibfnamefont {Alessandro}\
  \bibnamefont {Mirizzi}}, \bibinfo {author} {\bibfnamefont {Gianpiero}\
  \bibnamefont {Mangano}}, \ and\ \bibinfo {author} {\bibfnamefont {Ninetta}\
  \bibnamefont {Saviano}},\ }\bibfield  {title} {\enquote {\bibinfo {title}
  {{Self-induced flavor instabilities of a dense neutrino stream in a
  two-dimensional model}},}\ }\href {\doibase 10.1103/PhysRevD.92.021702}
  {\bibfield  {journal} {\bibinfo  {journal} {Phys. Rev.}\ }\textbf {\bibinfo
  {volume} {D92}},\ \bibinfo {pages} {021702} (\bibinfo {year} {2015})},\
  \Eprint {http://arxiv.org/abs/1503.03485} {arXiv:1503.03485 [hep-ph]}
  \BibitemShut {NoStop}%
\bibitem [{\citenamefont {Mirizzi}(2015)}]{Mirizzi:2015hwa}%
  \BibitemOpen
  \bibfield  {author} {\bibinfo {author} {\bibfnamefont {Alessandro}\
  \bibnamefont {Mirizzi}},\ }\bibfield  {title} {\enquote {\bibinfo {title}
  {{Breaking the symmetries in self-induced flavor conversions of neutrino
  beams from a ring}},}\ }\href {\doibase 10.1103/PhysRevD.92.105020}
  {\bibfield  {journal} {\bibinfo  {journal} {Phys. Rev.}\ }\textbf {\bibinfo
  {volume} {D92}},\ \bibinfo {pages} {105020} (\bibinfo {year} {2015})},\
  \Eprint {http://arxiv.org/abs/1506.06805} {arXiv:1506.06805 [hep-ph]}
  \BibitemShut {NoStop}%
\bibitem [{\citenamefont {Capozzi}\ \emph {et~al.}(2016)\citenamefont
  {Capozzi}, \citenamefont {Dasgupta},\ and\ \citenamefont
  {Mirizzi}}]{Capozzi:2016oyk}%
  \BibitemOpen
  \bibfield  {author} {\bibinfo {author} {\bibfnamefont {Francesco}\
  \bibnamefont {Capozzi}}, \bibinfo {author} {\bibfnamefont {Basudeb}\
  \bibnamefont {Dasgupta}}, \ and\ \bibinfo {author} {\bibfnamefont
  {Alessandro}\ \bibnamefont {Mirizzi}},\ }\bibfield  {title} {\enquote
  {\bibinfo {title} {{Self-induced temporal instability from a neutrino
  antenna}},}\ }\href {\doibase 10.1088/1475-7516/2016/04/043} {\bibfield
  {journal} {\bibinfo  {journal} {JCAP}\ }\textbf {\bibinfo {volume} {1604}},\
  \bibinfo {pages} {043} (\bibinfo {year} {2016})},\ \Eprint
  {http://arxiv.org/abs/1603.03288} {arXiv:1603.03288 [hep-ph]} \BibitemShut
  {NoStop}%
\bibitem [{\citenamefont {Duan}(2015)}]{Duan:2015cqa}%
  \BibitemOpen
  \bibfield  {author} {\bibinfo {author} {\bibfnamefont {Huaiyu}\ \bibnamefont
  {Duan}},\ }\bibfield  {title} {\enquote {\bibinfo {title} {{Collective
  neutrino oscillations and spontaneous symmetry breaking}},}\ }\href {\doibase
  10.1142/S0218301315410086} {\bibfield  {journal} {\bibinfo  {journal} {Int.
  J. Mod. Phys.}\ }\textbf {\bibinfo {volume} {E24}},\ \bibinfo {pages}
  {1541008} (\bibinfo {year} {2015})},\ \Eprint
  {http://arxiv.org/abs/1506.08629} {arXiv:1506.08629 [hep-ph]} \BibitemShut
  {NoStop}%
\bibitem [{\citenamefont {Abbar}\ and\ \citenamefont
  {Duan}(2015)}]{Abbar:2015fwa}%
  \BibitemOpen
  \bibfield  {author} {\bibinfo {author} {\bibfnamefont {Sajad}\ \bibnamefont
  {Abbar}}\ and\ \bibinfo {author} {\bibfnamefont {Huaiyu}\ \bibnamefont
  {Duan}},\ }\bibfield  {title} {\enquote {\bibinfo {title} {{Neutrino flavor
  instabilities in a time-dependent supernova model}},}\ }\href {\doibase
  10.1016/j.physletb.2015.10.019} {\bibfield  {journal} {\bibinfo  {journal}
  {Phys. Lett.}\ }\textbf {\bibinfo {volume} {B751}},\ \bibinfo {pages}
  {43--47} (\bibinfo {year} {2015})},\ \Eprint
  {http://arxiv.org/abs/1509.01538} {arXiv:1509.01538 [astro-ph.HE]}
  \BibitemShut {NoStop}%
\bibitem [{\citenamefont {Dasgupta}\ and\ \citenamefont
  {Mirizzi}(2015)}]{Dasgupta:2015iia}%
  \BibitemOpen
  \bibfield  {author} {\bibinfo {author} {\bibfnamefont {Basudeb}\ \bibnamefont
  {Dasgupta}}\ and\ \bibinfo {author} {\bibfnamefont {Alessandro}\ \bibnamefont
  {Mirizzi}},\ }\bibfield  {title} {\enquote {\bibinfo {title} {{Temporal
  Instability Enables Neutrino Flavor Conversions Deep Inside Supernovae}},}\
  }\href {\doibase 10.1103/PhysRevD.92.125030} {\bibfield  {journal} {\bibinfo
  {journal} {Phys. Rev.}\ }\textbf {\bibinfo {volume} {D92}},\ \bibinfo {pages}
  {125030} (\bibinfo {year} {2015})},\ \Eprint
  {http://arxiv.org/abs/1509.03171} {arXiv:1509.03171 [hep-ph]} \BibitemShut
  {NoStop}%
\bibitem [{\citenamefont {Sawyer}(2016)}]{Sawyer:2015dsa}%
  \BibitemOpen
  \bibfield  {author} {\bibinfo {author} {\bibfnamefont {R.~F.}\ \bibnamefont
  {Sawyer}},\ }\bibfield  {title} {\enquote {\bibinfo {title} {{Neutrino cloud
  instabilities just above the neutrino sphere of a supernova}},}\ }\href
  {\doibase 10.1103/PhysRevLett.116.081101} {\bibfield  {journal} {\bibinfo
  {journal} {Phys. Rev. Lett.}\ }\textbf {\bibinfo {volume} {116}},\ \bibinfo
  {pages} {081101} (\bibinfo {year} {2016})},\ \Eprint
  {http://arxiv.org/abs/1509.03323} {arXiv:1509.03323 [astro-ph.HE]}
  \BibitemShut {NoStop}%
\bibitem [{\citenamefont {Chakraborty}\ \emph
  {et~al.}(2016{\natexlab{b}})\citenamefont {Chakraborty}, \citenamefont
  {Hansen}, \citenamefont {Izaguirre},\ and\ \citenamefont
  {Raffelt}}]{Chakraborty:2016lct}%
  \BibitemOpen
  \bibfield  {author} {\bibinfo {author} {\bibfnamefont {Sovan}\ \bibnamefont
  {Chakraborty}}, \bibinfo {author} {\bibfnamefont {Rasmus~Sloth}\ \bibnamefont
  {Hansen}}, \bibinfo {author} {\bibfnamefont {Ignacio}\ \bibnamefont
  {Izaguirre}}, \ and\ \bibinfo {author} {\bibfnamefont {Georg}\ \bibnamefont
  {Raffelt}},\ }\bibfield  {title} {\enquote {\bibinfo {title} {{Self-induced
  neutrino flavor conversion without flavor mixing}},}\ }\href {\doibase
  10.1088/1475-7516/2016/03/042} {\bibfield  {journal} {\bibinfo  {journal}
  {JCAP}\ }\textbf {\bibinfo {volume} {1603}},\ \bibinfo {pages} {042}
  (\bibinfo {year} {2016}{\natexlab{b}})},\ \Eprint
  {http://arxiv.org/abs/1602.00698} {arXiv:1602.00698 [hep-ph]} \BibitemShut
  {NoStop}%
\bibitem [{\citenamefont {Cirigliano}\ \emph {et~al.}(2017)\citenamefont
  {Cirigliano}, \citenamefont {Paris},\ and\ \citenamefont
  {Shalgar}}]{Cirigliano:2017hmk}%
  \BibitemOpen
  \bibfield  {author} {\bibinfo {author} {\bibfnamefont {Vincenzo}\
  \bibnamefont {Cirigliano}}, \bibinfo {author} {\bibfnamefont {Mark~W.}\
  \bibnamefont {Paris}}, \ and\ \bibinfo {author} {\bibfnamefont {Shashank}\
  \bibnamefont {Shalgar}},\ }\bibfield  {title} {\enquote {\bibinfo {title}
  {{Effect of collisions on neutrino flavor inhomogeneity in a dense neutrino
  gas}},}\ }\href {\doibase 10.1016/j.physletb.2017.09.039} {\bibfield
  {journal} {\bibinfo  {journal} {Phys. Lett.}\ }\textbf {\bibinfo {volume}
  {B774}},\ \bibinfo {pages} {258--267} (\bibinfo {year} {2017})},\ \Eprint
  {http://arxiv.org/abs/1706.07052} {arXiv:1706.07052 [hep-ph]} \BibitemShut
  {NoStop}%
\bibitem [{\citenamefont {Capozzi}\ \emph {et~al.}(2018)\citenamefont
  {Capozzi}, \citenamefont {Dasgupta}, \citenamefont {Mirizzi}, \citenamefont
  {Sen},\ and\ \citenamefont {Sigl}}]{Capozzi:2018clo}%
  \BibitemOpen
  \bibfield  {author} {\bibinfo {author} {\bibfnamefont {Francesco}\
  \bibnamefont {Capozzi}}, \bibinfo {author} {\bibfnamefont {Basudeb}\
  \bibnamefont {Dasgupta}}, \bibinfo {author} {\bibfnamefont {Alessandro}\
  \bibnamefont {Mirizzi}}, \bibinfo {author} {\bibfnamefont {Manibrata}\
  \bibnamefont {Sen}}, \ and\ \bibinfo {author} {\bibfnamefont {Günter}\
  \bibnamefont {Sigl}},\ }\bibfield  {title} {\enquote {\bibinfo {title}
  {{Collisional triggering of fast flavor conversions of supernova
  neutrinos}},}\ }\href@noop {} {\  (\bibinfo {year} {2018})},\ \Eprint
  {http://arxiv.org/abs/1808.06618} {arXiv:1808.06618 [hep-ph]} \BibitemShut
  {NoStop}%
\bibitem [{\citenamefont {Chakraborty}\ \emph
  {et~al.}(2016{\natexlab{c}})\citenamefont {Chakraborty}, \citenamefont
  {Hansen}, \citenamefont {Izaguirre},\ and\ \citenamefont
  {Raffelt}}]{Chakraborty:2016yeg}%
  \BibitemOpen
  \bibfield  {author} {\bibinfo {author} {\bibfnamefont {Sovan}\ \bibnamefont
  {Chakraborty}}, \bibinfo {author} {\bibfnamefont {Rasmus}\ \bibnamefont
  {Hansen}}, \bibinfo {author} {\bibfnamefont {Ignacio}\ \bibnamefont
  {Izaguirre}}, \ and\ \bibinfo {author} {\bibfnamefont {Georg}\ \bibnamefont
  {Raffelt}},\ }\bibfield  {title} {\enquote {\bibinfo {title} {{Collective
  neutrino flavor conversion: Recent developments}},}\ }\href {\doibase
  10.1016/j.nuclphysb.2016.02.012} {\bibfield  {journal} {\bibinfo  {journal}
  {Nucl. Phys.}\ }\textbf {\bibinfo {volume} {B908}},\ \bibinfo {pages}
  {366--381} (\bibinfo {year} {2016}{\natexlab{c}})},\ \Eprint
  {http://arxiv.org/abs/1602.02766} {arXiv:1602.02766 [hep-ph]} \BibitemShut
  {NoStop}%
\bibitem [{\citenamefont {Sigl}\ and\ \citenamefont
  {Raffelt}(1993)}]{Sigl:1992fn}%
  \BibitemOpen
  \bibfield  {author} {\bibinfo {author} {\bibfnamefont {G.}~\bibnamefont
  {Sigl}}\ and\ \bibinfo {author} {\bibfnamefont {G.}~\bibnamefont {Raffelt}},\
  }\bibfield  {title} {\enquote {\bibinfo {title} {{General kinetic description
  of relativistic mixed neutrinos}},}\ }\href {\doibase
  10.1016/0550-3213(93)90175-O} {\bibfield  {journal} {\bibinfo  {journal}
  {Nucl. Phys.}\ }\textbf {\bibinfo {volume} {B406}},\ \bibinfo {pages}
  {423--451} (\bibinfo {year} {1993})}\BibitemShut {NoStop}%
\bibitem [{\citenamefont {Duan}\ \emph
  {et~al.}(2006{\natexlab{c}})\citenamefont {Duan}, \citenamefont {Fuller},\
  and\ \citenamefont {Qian}}]{Duan:2005cp}%
  \BibitemOpen
  \bibfield  {author} {\bibinfo {author} {\bibfnamefont {Huaiyu}\ \bibnamefont
  {Duan}}, \bibinfo {author} {\bibfnamefont {George~M.}\ \bibnamefont
  {Fuller}}, \ and\ \bibinfo {author} {\bibfnamefont {Yong-Zhong}\ \bibnamefont
  {Qian}},\ }\bibfield  {title} {\enquote {\bibinfo {title} {{Collective
  neutrino flavor transformation in supernovae}},}\ }\href {\doibase
  10.1103/PhysRevD.74.123004} {\bibfield  {journal} {\bibinfo  {journal} {Phys.
  Rev.}\ }\textbf {\bibinfo {volume} {D74}},\ \bibinfo {pages} {123004}
  (\bibinfo {year} {2006}{\natexlab{c}})},\ \Eprint
  {http://arxiv.org/abs/astro-ph/0511275} {arXiv:astro-ph/0511275 [astro-ph]}
  \BibitemShut {NoStop}%
\bibitem [{\citenamefont {Kostelecky}\ and\ \citenamefont
  {Samuel}(1995)}]{Kostelecky:1994dt}%
  \BibitemOpen
  \bibfield  {author} {\bibinfo {author} {\bibfnamefont {V.~Alan}\ \bibnamefont
  {Kostelecky}}\ and\ \bibinfo {author} {\bibfnamefont {Stuart}\ \bibnamefont
  {Samuel}},\ }\bibfield  {title} {\enquote {\bibinfo {title} {{Selfmaintained
  coherent oscillations in dense neutrino gases}},}\ }\href {\doibase
  10.1103/PhysRevD.52.621} {\bibfield  {journal} {\bibinfo  {journal} {Phys.
  Rev.}\ }\textbf {\bibinfo {volume} {D52}},\ \bibinfo {pages} {621--627}
  (\bibinfo {year} {1995})},\ \Eprint {http://arxiv.org/abs/hep-ph/9506262}
  {arXiv:hep-ph/9506262 [hep-ph]} \BibitemShut {NoStop}%
\bibitem [{\citenamefont {Duan}\ \emph {et~al.}(2007)\citenamefont {Duan},
  \citenamefont {Fuller}, \citenamefont {Carlson},\ and\ \citenamefont
  {Qian}}]{Duan:2007mv}%
  \BibitemOpen
  \bibfield  {author} {\bibinfo {author} {\bibfnamefont {Huaiyu}\ \bibnamefont
  {Duan}}, \bibinfo {author} {\bibfnamefont {George~M.}\ \bibnamefont
  {Fuller}}, \bibinfo {author} {\bibfnamefont {J.}~\bibnamefont {Carlson}}, \
  and\ \bibinfo {author} {\bibfnamefont {Yong-Zhong}\ \bibnamefont {Qian}},\
  }\bibfield  {title} {\enquote {\bibinfo {title} {{Analysis of Collective
  Neutrino Flavor Transformation in Supernovae}},}\ }\href {\doibase
  10.1103/PhysRevD.75.125005} {\bibfield  {journal} {\bibinfo  {journal} {Phys.
  Rev.}\ }\textbf {\bibinfo {volume} {D75}},\ \bibinfo {pages} {125005}
  (\bibinfo {year} {2007})},\ \Eprint {http://arxiv.org/abs/astro-ph/0703776}
  {arXiv:astro-ph/0703776 [astro-ph]} \BibitemShut {NoStop}%
\bibitem [{\citenamefont {Press}\ \emph {et~al.}(2002)\citenamefont {Press},
  \citenamefont {Teukolsky}, \citenamefont {Vetterling},\ and\ \citenamefont
  {Flannery}}]{NR2002}%
  \BibitemOpen
  \bibfield  {author} {\bibinfo {author} {\bibfnamefont {William~H.}\
  \bibnamefont {Press}}, \bibinfo {author} {\bibfnamefont {Saul~A.}\
  \bibnamefont {Teukolsky}}, \bibinfo {author} {\bibfnamefont {William~T.}\
  \bibnamefont {Vetterling}}, \ and\ \bibinfo {author} {\bibfnamefont
  {Brian~P.}\ \bibnamefont {Flannery}},\ }\href@noop {} {\emph {\bibinfo
  {title} {Numerical Recipes in C++: The Art of Scientific Computing}}},\
  \bibinfo {edition} {2nd}\ ed.\ (\bibinfo  {publisher} {Cambridge University
  Press},\ \bibinfo {address} {Cambridge},\ \bibinfo {year} {2002})\BibitemShut
  {NoStop}%
\bibitem [{\citenamefont {Raffelt}\ and\ \citenamefont
  {Sigl}(2007)}]{Raffelt:2007yz}%
  \BibitemOpen
  \bibfield  {author} {\bibinfo {author} {\bibfnamefont {G.~G.}\ \bibnamefont
  {Raffelt}}\ and\ \bibinfo {author} {\bibfnamefont {G.}~\bibnamefont {Sigl}},\
  }\bibfield  {title} {\enquote {\bibinfo {title} {{Self-induced decoherence in
  dense neutrino gases}},}\ }\href {\doibase 10.1103/PhysRevD.75.083002}
  {\bibfield  {journal} {\bibinfo  {journal} {Phys. Rev.}\ }\textbf {\bibinfo
  {volume} {D75}},\ \bibinfo {pages} {083002} (\bibinfo {year} {2007})},\
  \Eprint {http://arxiv.org/abs/hep-ph/0701182} {arXiv:hep-ph/0701182 [hep-ph]}
  \BibitemShut {NoStop}%
\end{thebibliography}%

\end{document}